\documentclass[aps,pre,twocolumn,10pt,showpacs,superscriptaddress]{revtex4-2}
\usepackage[english]{babel}
\usepackage[utf8x]{inputenc} 
\usepackage[T1]{fontenc}
\usepackage{amsmath}
\usepackage{graphicx}
\usepackage{color}
\usepackage{listings}
\usepackage{float}
\usepackage{verbatim}
\usepackage[colorlinks=true, linkcolor=blue, citecolor=blue, urlcolor=blue]{hyperref}
\begin{document}

\title{Thermalization in a nonlinear variant of the discrete nonlinear Schrödinger equation}

\author{Yagmur Kati}
\email{yagmur.kati@hu-berlin.de}
\affiliation{Department of Physics, Humboldt University of Berlin, Berlin, Germany}
\author{Aleksandra Maluckov}
\affiliation{Vinča Institute of Nuclear Sciences, National Institute of the Republic of Serbia, University of Belgrade, POB. 522, 11001 Belgrade, Serbia}
\author{Ana Mancic}
\affiliation{Faculty of Sciences and Mathematics, University of Niš, Višegradska 33, 18000 Niš, Serbia}
\author{Panayotis Kevrekidis}
\affiliation{Department of Mathematics and Statistics, University of Massachusetts Amherst, Amherst, Massachusetts, USA}
\affiliation{Department of Physics, University of Massachusetts Amherst, Amherst, Massachusetts, USA}
\begin{abstract} 
We study the thermalization properties of a fully nonlinear lattice model originally derived from the two-dimensional cubic defocusing nonlinear Schr{\"o}dinger equation (NLS) using analytical and numerical methods.
The model conserves both energy and norm, whose densities define a microcanonical energy-norm parameter space, while the nonlinear nearest-neighbor coupling is controlled by a parameter $D$.
Within this space, our analysis identifies broad parameter regimes in which the dynamics is ergodic not only within but also outside the standard Gibbs region, indicating the need for a modified statistical description.
At higher energies, the system instead exhibits long-lived compacton-mediated localization and signatures of weak nonergodicity, as evidenced by finite-time variances, excursion-time statistics, and probability distributions of local amplitudes.
We show that stronger coupling $D$ enhances fluctuations and accelerates
the crossover of the finite-time variance of the local norm density from an initial $\sim T^{-1/2}$ decay toward the faster $\sim T^{-1}$ decay characteristic of ergodic thermalization, where $T$ denotes the averaging time.
In the high-energy regime, weak coupling $D\leq 1$ favors persistent single-site compacton localization, whereas stronger coupling $D>1$ yields long-lived two-site localization.
Our results provide insights into the interplay between thermalization, localization, and non-Gibbs statistical behavior in genuinely nonlinear systems.
\end{abstract}

\maketitle

\section{Introduction}

Nonlinear dispersive lattice models have emerged
as an important modeling platform for a wide
variety of physical settings. These range from
optical waveguide arrays~\cite{LEDERER20081}
to mean-field atomic condensates in optical
lattices~\cite{RevModPhys.78.179}
and from granular crystals~\cite{Nester2001,yuli_book}
and cantilever arrays~\cite{cantilevers}
to nonlinear electrical circuits~\cite{remoissenet}
and biophysical settings such as the DNA double
strand~\cite{PeyBi,DP06}. This diverse set of
experimental applications and the 
associated theoretical and computational tools of this field have by now been summarized in 
numerous reviews~\cite{Aubry2006,Flach2008},
as well as books~\cite{remoissenet,DP06,Chong2018}.

Among the relevant models, arguably the most 
prototypical one, operating as an envelope model
broadly applicable (through appropriate reductions)
to most of the above applications, one can mention
the discrete nonlinear Schr{\"o}dinger (DNLS)
equation~\cite{chriseil,AblowitzPrinariTrubatch,kev09}.
The DNLS model combines the principal ingredients
of dispersive nearest-neighbor lattice dynamics
and onsite (cubic) nonlinearity and as such
is central to applications, e.g., in nonlinear
optics of waveguide arrays~\cite{LEDERER20081},
as well as to tunneling-coupled 
Bose-Einstein condensates at the wells of an 
optical lattice~\cite{RevModPhys.78.179} [a topic that remains
extremely active, including in experiments, to this day~\cite{cruickshank}].
The model supports a large variety of solitary
waves that have been extensively analyzed in
one spatial dimension, while more exotic
vortical states exist in higher dimensional
settings~\cite{kev09}.

The study of thermalization and localization in nonlinear 
dynamical lattices is crucial for understanding a wide range of physical phenomena, from energy transport in condensed matter to wave dynamics in optical lattices; after
all this was fundamental in the original query of E. Fermi
that led to the Fermi-Pasta-Ulam-Tsingou model and
the birth of nonlinear science~\cite{Fermi55,zabu}.
In the context of the more standard DNLS model,
this type of query has been considered in a 
substantial volume of publications following~\cite{Rasmussen00}. Nevertheless,
relevant studies focusing on ergodicity properties
and on energy localization, developing different
types of diagnostics as well as exploring 
a diverse array of systems, are still a highly
active area of research\cite{Iubini_2013,MithunDNLS18,MithunJJ19,Danieli19,Kati20,Kati21,Merab22},
including through the development of the so-called
``optical thermodynamics'' of weakly nonlinear lattices~\cite{WuHassanChristodoulides2019,EfremidisChristodoulides2022,MakrisWuJungChristodoulides2020,lian2026nonlinearopticalthermodynamicsvan} and its extension to strongly nonlinear regimes~ \cite{kabat2026opticalthermodynamicsweaknonlinearity}. The study of the related
possibility of reaching negative 
temperatures has recently been reviewed
in~\cite{BALDOVIN20211} and the associated interpretation of localization as
a form of condensation continues to be a topic of ongoing 
investigation~\cite{IubiniPoliti2025}.

Recently, a different class of DNLS models
has emerged in a completely different setting.
In the work of the so-called I-team
a reduction from the two-dimensional
defocusing cubic nonlinear Schrödinger equation
on the torus~\cite{Colliander10,Colliander13}
led to a lattice model intended as a toy
representation of turbulent cascades.
Indeed, in this model
the notion of a ``lattice node''
can be thought of as 
representing a group of wavenumbers in the Fourier
space. Thus, potential
mobility within this model could be thought of
as representing the transfer of energy 
across frequencies. Among the intriguing
features of this model one can list the
fact that it is genuinely nonlinear
(i.e., there is no linear dispersive part)
and that it can thus support compactly supported
nonlinear excitations~\cite{Kevrekidis23}.
The intrinsic interest within this type
of setting led to studies of its numerical
features~\cite{Colliander13}, of numerical 
schemes to integrate such models~\cite{gideon2} and of its
connections to solutions of a discrete Burgers
equation~\cite{Her}.

The present work lies at the nexus of all these
intriguing, ongoing themes of research. 
It extends the study of the thermodynamics of the DNLS ~\cite{Rasmussen00,MithunDNLS18} and of the recent attempt ~\cite{PhysRevE.103.032211} to go beyond the DNLS and towards approaching the integrable Ablowitz–Ladik analogue thereof~\cite{AblowitzPrinariTrubatch}, with these works involving subsets of the present authors, through the thermodynamics of the so-called Salerno model~\cite{salerno1992quantum}.
It connects this ``thermodynamics of nonlinear
lattice models'' thread of research 
via the key transfer integral methodology of~\cite{PhysRevB.6.3409} 
with the setting of genuinely nonlinear
models such as the one of~\cite{Colliander10},
which no longer features solitary waves
and regular discrete breathers, but rather
focuses the energy on discrete compacton
states~\cite{vvk1,vvk3}. This is also emerging as a challenging frontier of statistical mechanics research efforts on nonlinear lattices, as is evidenced by the recent work of~\cite{MarzuolaMattingly2025}, aiming to construct the invariant measure of this lattice through a characterization of its energy minimizers.

The scope of the present study is to advance the current understanding of thermalization in nonlinear lattice models by providing insights into the intricate interplay between (genuine) nonlinearity, (compact) localization, and ergodicity as they emerge in the intriguing toy model of the I-team~\cite{Colliander10}. 
Our presentation is structured as follows. In section \ref{sec:model} we introduce the model of interest, outline the relevant conservation laws, and describe the thermodynamic formalism based on the transfer integral. 
We use section \ref{sec:method} to develop a series of tools that will be used in the computational analysis of the toy model, including finite time (mass or $l^2$ norm) averages, probability density functions (PDFs) of so-called excursion times, {and the probability distribution $p(A)$ of local amplitudes}, i.e., some of the tools developed recently to explore the dynamics and thermodynamics of such lattice systems~\cite{MithunDNLS18,MithunJJ19,Danieli19,Merab22,Rasmussen00,PhysRevE.103.032211}. 
In section \ref{sec:discussion} we present a discussion of our findings. Finally, in section \ref{sec:conclusion}, we summarize our conclusions, and offer an outlook towards possible future studies.

\section{Toy Model System \& Thermodynamic Formalism} \label{sec:model}

The toy model of interest herein is derived from the nonlinear Schrödinger equation (NLS) on a two-dimensional toroidal domain with periodic boundary conditions. This setup facilitates the analysis of wave propagation and interaction effects. Colliander \textit{et al.} \cite{Colliander13} reduced the NLS to the toy model, described by the equations of motion:
\begin{equation}
 i \dot{b_\ell} = |b_\ell|^2 b_\ell-D  {b_\ell}^* (b_{\ell-1}^2 + b_{\ell+1}^2 )\label{eom}
\end{equation}
with boundary conditions $b_0 = b_{N+1} = 0$, where the overdot represents the time derivative, $\ell$ is the site index, and $D$ is the parameter controlling the interaction between nearest neighbors. The equations of motion, given by
\begin{equation}
    i\dot{b}_\ell=2\frac{\partial H}{\partial {b_\ell}^*}
\end{equation}
define the system's dynamics, where $H$ is the Hamiltonian:
\begin{equation}\label{hamiltoniantoy}
    H = \sum_{\ell=1}^N \frac{1}{4} |b_\ell|^4 - \frac{D}{2} \text{Re}({{b_\ell^*}}^2 {b_{\ell-1}}^2).
\end{equation} 
The dynamics of 
Eq.~(\ref{eom}) conserve both the total energy $H$ and the total norm $A=\sum_\ell |b_\ell|^2$. The two conserved quantities $A$ and $H$ give us a two-dimensional parameter space $(a,h)$ to define our initial states, where $a=A/N$ is the norm density and $h=H/N$ is the energy density of the system.

\begin{figure}[ht]
\centering
\includegraphics[width=0.49\textwidth]{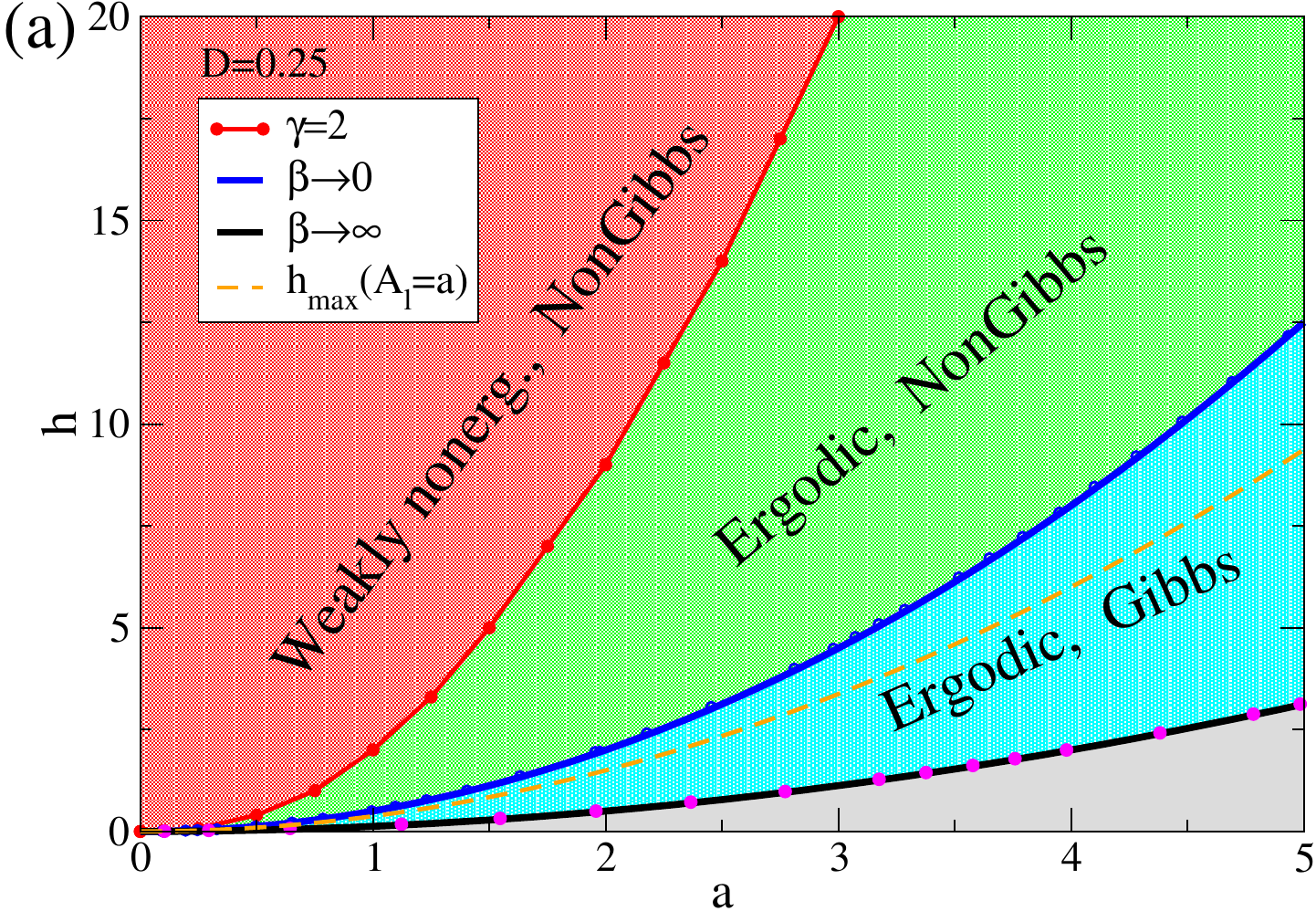}
\includegraphics[width=0.49\textwidth]{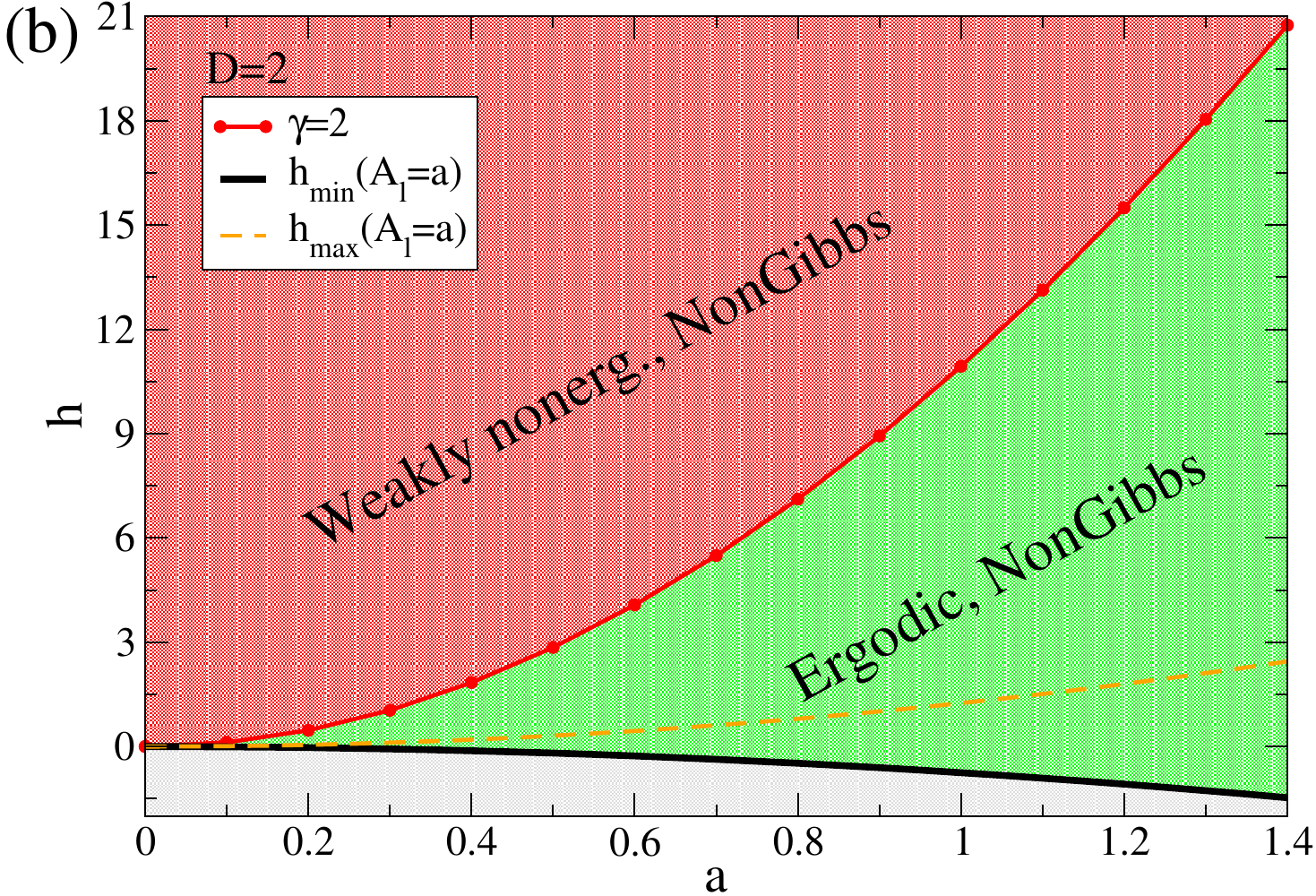}  
\caption{Phase diagram for the toy model: (a) $D=0.25$ and (b) $D=2$.
In (a), the $\beta=0$ and $\beta \rightarrow \infty$ curves correspond to $h=a^2/2$ (blue line, TIO blue circles) and $h=a^2/8$ (black line, TIO magenta circles).
Red circles denote initial states with excursion-time exponent $\gamma=2$.
The orange dashed line is the homogeneous-norm upper bound $h_{\max}(A_\ell=a)=\tfrac{a^2}{4}(1+2D)$, and the black solid line the corresponding lower bound $h_{\min}(A_\ell=a)=\tfrac{a^2}{4}(1-2D)$; for $D=0.25$ this coincides with $h(\beta \rightarrow \infty)$.
For $D>0.5$ [panel (b)], heterogeneous norm distributions allow energies unbounded from below in the thermodynamic limit, so $h_{\min}(A_\ell=a)$ is not the ground state.
The TIO approach is applicable only within the cyan region of panel (a).}
    \label{fig:phase}
\end{figure}

To examine the system’s statistical properties, we define and determine the classical grand-canonical partition function $Z$ following~\cite{Rasmussen00}; see also the
pioneering work of~\cite{PhysRevB.6.3409}. We start with a canonical transformation, $b_\ell= \sqrt{A_\ell} e^{i \phi_\ell}$, which yields the Hamiltonian in the form: 

\begin{equation}\label{eq:H_A_ell}
   H=\sum_\ell^N \frac{1}{4} |A_\ell|^2 -\frac{D}{2}{A_\ell A_{\ell-1}} \cos(2\phi_\ell-2\phi_{\ell-1}).
\end{equation}
This leads to the expression for the partition function in the grand-canonical
ensemble:
\begin{equation}\label{Zgeneral}
    Z=\int_0^\infty \int_0^{2\pi} \prod_{\ell=1}^N d\phi_\ell dA_\ell e^{-\beta (H+\alpha A)},
\end{equation}
where $\beta$ is the inverse temperature and the Lagrange multiplier $\alpha$ serves as a chemical potential associated with the conservation of $A$. Integrating over the phase variable $\phi_\ell$, and symmetrizing the partition function we obtain:
\begin{align}\label{partitionfunction}
&Z = (2\pi)^N \int_0^\infty \prod_\ell dA_\ell \, I_0\left[\frac{\beta D}{2}A_\ell A_{\ell-1}\right] \nonumber\\
&\times \exp\left[ \sum_\ell -\frac{\beta}{8}(A_\ell^2 + A_{\ell-1}^2) - \frac{\beta \alpha}{2}(A_\ell + A_{\ell-1}) \right].
\end{align}
In the thermodynamic limit  ($N\rightarrow\infty$), the integral in $Z$ can be evaluated exactly by using the eigenvalues and eigenfunctions of the transfer integral operator  (TIO) \cite{Rasmussen00,PhysRevB.6.3409,PhysRevB.11.3535}, a fundamental technique
enabling such analysis for one-dimensional systems. In this limit, the partition function is approximated as $Z\approx (2\pi\lambda_0)^N$, where $\lambda_0$ is the largest eigenvalue of the TIO [with exponentially small corrections associated with
$\propto (\lambda_1/\lambda_0)^N$ where $\lambda_1$ is the next eigenvalue of the relevant
operator], allowing us to compute the integral as:
\begin{equation}\label{eq:kernelintegral}
    \int_0^\infty dA_\ell \kappa(A_\ell, A_{\ell-1}) y(A_\ell)=\lambda y(A_{\ell-1}),
\end{equation}
where the kernel matrix is
\begin{equation}\label{eq:kernel}
\kappa(x,z)=I_0\left[\frac{D}{2}\beta x z\right] \exp{\left[ \sum_\ell -\frac{\beta}{8}(x^2+z^2) -\frac{\beta\alpha}{2}(x+z)\right]}. 
\end{equation}
Here $x$ and $z$ are continuous amplitude variables corresponding to two neighboring sites, $x\equiv A_\ell$ and $z\equiv A_{\ell-1}$, which are integrated from 0 to $\infty$. 
In the context of Gibbs thermodynamics, a well-defined temperature requires a non-divergent kernel matrix as $\beta \rightarrow 0$ and $\beta \rightarrow \infty$. 

For values $D > 0.5$, however, the kernel diverges, rendering the transfer integral method {\it inapplicable}. As  $D$  approaches  0.5, the kernel assumes the asymptotic form:
\begin{equation} \label{eq:kappaxx}
    \kappa(x,x) = \frac{\exp{\left[-\frac{\beta x^2}{4}(1 - 2D) - \beta \alpha x\right]}}{\sqrt{\pi D \beta x^2}}.
\end{equation}
For $D>0.5$ (and $\beta>0$), the quadratic term changes sign and the exponent grows as $+\frac{\beta}{4}(2D-1)x^2$, implying an exponential divergence of $\kappa(x,x)$ at large $x$. The case $D=0.5$ is marginal: the quadratic term vanishes and $\kappa(x,x)\propto e^{-\beta\alpha x}/x$, so convergence requires $\beta\alpha>0$ (in particular, $\alpha>0$ for $\beta>0$). Therefore, kernel convergence (and applicability of the transfer-integral method) is ensured for $D<0.5$, and for $D=0.5$ provided $\beta\alpha>0$.

Since a valid Gibbs temperature curve in the parameter space $(a,h)$ (see Fig.\ref{fig:phase}) is not generally possible for $D > 0.5$, we 
at first consider the case $D=0.25$ as representative (of the regime where
thermalization ---and associated computation via TIO--- is possible) to 
start the investigation of the system relaxation. 
With the transfer integral method, an initial state in the parameter space $(a,h)$ can correspond to the Gibbs parameters ($\beta,\alpha$) 
\begin{equation} \label{eq:a}
    a=-\frac{1}{\beta\lambda_0}\frac{\partial\lambda_0}{\partial\alpha}, \quad h=-\frac{1}{\lambda_0}\frac{\partial\lambda_0}{\partial\beta}-\alpha a.
\end{equation}
These equations are expected to provide a one-to-one mapping between the dynamical
quantities of a given initialization $(a,h)$ and the corresponding thermodynamic
ones ($\beta,\alpha$) of the thermalized asymptotic state, when thermalization
is achieved.

We define the finite temperature region in the parameter space $(a,h)$ to investigate the Gibbs regime, where the system thermalizes and its statistics is well-defined.  By applying  Eq.~(\ref{eq:a}), we derive the infinite-temperature line ($\beta=0$) as $a=\frac{1}{\alpha\beta}$ and  $h= \frac{1}{2\alpha^2\beta^2} =  \frac{a^2}{2}$ (see Appendix~\ref{sec:temperaturecurves}). 
{
Parameter values outside the normalizable standard Gibbs construction will be referred to as non-Gibbs. Thus, for $D\leq0.5$, the non-Gibbs regime lies beyond the $\beta=0$ line, whereas for $D>0.5$, the kernel divergence discussed above prevents a Gibbs region altogether.} 
We can reach the ground state with a plane wave $b_\ell=\sqrt{a} e^{i\pi \ell/2}$ \cite{Rasmussen00} that suggests a constant norm density distribution $A_\ell=a$. Inserting it into Eq.(\ref{eq:H_A_ell}) gives  
\begin{equation}\label{eq:H_A_ell2}
    h= \frac{1}{4} a^2 -\frac{D}{2}a^2 \cos(2\Delta\phi), 
\end{equation}
where the phase difference $\Delta\phi=0$ gives the zero temperature curve $h=\frac{1-2D}{4} a^2$ for $D<0.5$. Similarly, the maximum energy density that would be reached with a constant norm is $h=\frac{1+2D}{4} a^2$, setting $\Delta\phi=\pi/2$ for all $\phi_\ell$. For our numerical results, we define the initial condition as follows.
\begin{equation}\label{eq:b_ell}
b_\ell = \sqrt{A_\ell} \, e^{i\phi_\ell},   
\end{equation}
where $ A_\ell $ represents the amplitude and $ \phi_\ell $ the phase. 
In all simulations presented in this paper, we set the number of sites to $N=1000$. 
When it is possible to define the initial condition by a constant norm i.e. $\frac{1-2D}{4} a^2<h<\frac{1+2D}{4} a^2$, we simply set $A_\ell=a + \delta_\ell$ and use a fixed phase difference with a small perturbation $\phi_\ell=\ell \Delta\phi + \delta'_\ell$, where
\begin{equation}
    \Delta\phi=\frac{1}{2} \arccos{\left[\frac{1}{2D}-\frac{2h}{Da^2}\right]}.
\end{equation}
Here, $\delta_\ell$ and $\delta'_\ell$ are two independent sets of random numbers which are uniformly distributed over $[-0.001, 0.001]$. We introduce them as a small disturbance to both $A_\ell$ and $\phi_\ell$ for the numerical simulations.

Note that for $D= 0.5$, the infinite-temperature line {$h=a^2/2$ coincides with}
the maximum energy density line reached by a constant norm, that is, $h_{\text{max}}(A_\ell=a)=\frac{(1+2D)a^2}{4}$.
{In Fig.\ref{fig:phase}, only the region on and below the orange $h_{\text{max}}$ line can be accessed with a constant amplitude distribution. Regions above this line, including the non-Gibbs regimes in Figs.~\ref{fig:phase}(a) and \ref{fig:phase}(b) and a significant Gibbs region in Fig.~\ref{fig:phase}(a), require nonconstant norm-density distributions.} 
We can prepare the associated initial states as
\begin{equation}\label{eq:ic2}
b_\ell=\sqrt{A_\ell} e^{i \pi \ell/2}, \quad A_\ell = a +  \, \epsilon_\ell^c \, \exp\left[ x \eta_\ell  \right],
\end{equation}
where $\epsilon_\ell$ and  $\eta_\ell$  are independent random variables, with  $\epsilon_\ell$  uniformly distributed in $[0, 1]$ and  $\eta_\ell$  uniformly distributed in [-1, 1]. $x$ and $c$ are non-negative real numbers that control the energy density. Here we set $\Delta\phi$ to $\pi/2$, and  $a$ to the desired norm density, which is 0.25 for most of the plots shown in this paper.

As we just mentioned, a random seed is used for $\epsilon_\ell, \eta_\ell$ for each realization of the initial states with a high $H/A^2$. Therefore, one may consider that an initial state with strongly excited two or more adjacent sites is also possible.
Hence, we additionally prepare an atypical but probable initial state that is not used for average statistics such as the ones
explained in the following sections \ref{sec:finitetimeaverages} and \ref{sec:pdf}, but to test the existence and survival of several site compactons (dynamically relevant
to this model in line
also with the discussion of~\cite{Kevrekidis23}), as described in section \ref{sec:compacton}. 

These compactly supported states are constructed as follows. 
After we prepare the high-energy initial state as in Eq.(\ref{eq:ic2}), 
we set $A_0=\text{max}(A_\ell)/m$, and then additionally excite $m$ adjacent sites in the middle of the lattice with $b_S= \sqrt{A_0}e^{i\phi_S}$ where $S=\{\ell_0,\ldots,\ell_0+m-1\}$. We set $\phi_S= S \Delta\phi=S \pi/2$ unless it is a test run for the low-energy in-phase structures, where $\Delta\phi$ is set to zero. For $m=2$, we choose $\ell_0=N/2$, and for $m=3$, $\ell_0=N/2-1$. The initial state for the compacton tests can be described as follows
\begin{align}
&b_\ell=\sqrt{A_\ell} e^{i \pi \ell/2}, \quad A_\ell = a +  \, \epsilon_\ell^c \, \exp\left[ x \eta_\ell  \right],\,\,\, A_0=\frac{\text{max}(A_\ell)}{m} \nonumber\\
&b_S= \sqrt{A_0}e^{i \phi_S}, \,\, S=\{\ell_0,\ldots,\ell_0+m-1\}, \,\,\ell\notin S. \label{eq:compact}
\end{align}

In this manuscript, for the initial-states defined by Eq.~(\ref{eq:compact}),
we will share norm density pictures only for $m=2$ and $m=3$. Note that 
we also tested several adjacent excited-site cases with $m=4,\dots,14$; however, these multi-site structures ($m>3$) did not survive to later stages of the evolution, nor did they form spontaneously later. Instead, they rapidly converged to one, two or three site moving structures (the transition is not shown here, see Fig.~\ref{fig:energymaximizer} for analytical reasoning).

For all initial conditions described above, to maintain the desired average norm density $a$, we normalize the lattice by first calculating the current total norm $A_c$, then rescaling each $b_\ell$ by $b_\ell \rightarrow b_\ell \sqrt{a N / A_c}$ such that $\sum_{\ell=1}^N |b_\ell|^2 = a N$.

Fig.~\ref{fig:phase} shows the phase diagrams of energy density $h$ versus norm density $a$ for $D=0.25$ and $D=2$. According to equilibrium statistical mechanics \cite{RevModPhys.92.041002,adler2004quantum} the Hamiltonian has to be bounded from below to have a convergent partition function. 
In both panels, the solid black line represents the minimum energy reached by a constant norm density, $h_{\mathrm{min}}(a_\ell=a) = \frac{a^2}{4}(1-2D)$ (see Eq.~(\ref{eq:minenergy}) for details). 
For $D = 0.25 < 0.5$, this curve is convex and corresponds to the true ground-state line, meaning it is size-independent and the region below is inaccessible to any initial state.  For $D = 2$, the same equation becomes concave and no longer represents the ground state. In this case, by heterogeneously redistributing the norm, the energy can be reduced up to $h_{\min}(a) = -7 a^{2}N/44$ (see Eq.~(\ref{eq:H3in}) below). The latter is the minimum for $D=2$ \cite{Kevrekidis23} that is reached by only exciting three sites with $A_\ell=A_{\ell+2}=3, A_{\ell+1}=5$ (for $D=2$, the minimum energy is achieved by exciting only three neighboring sites with a 3-5-3 proportionality of the respective square moduli) and normalizing the lattice to have the desired $a$, as described in the previous paragraph. In the thermodynamic limit $N \to \infty$, the minimum energy density becomes unbounded from below, and no global ground-state line exists. Hence, note that the gray area in Fig~\ref{fig:phase}(b) is accessible and non-Gibbs thermalizable in its nature, while the gray area in Fig.\ref{fig:phase}(a) is inaccessible. It is relevant to mention here that, {in line with~\cite{MithunDNLS18}, for $D < 0.5$, where the model behavior is most analogous to the standard DNLS, one may similarly expect} a regime of $(a,h)$ parameters beyond {$\beta=0$ that is non-Gibbs yet still ergodic, and at even higher energies, a weakly nonergodic (and non-Gibbs) regime where persistent norm localization strongly delays global equilibration.}

Our main conclusions from the results of Fig.~\ref{fig:phase}  are then the following. 
In the regime where TIO is applicable ($D \leq 0.5$), 
we expect the system to thermalize, i.e., to reach thermodynamic equilibrium with a given temperature
(between $0$ and $\infty$) and chemical potential for a given pair of $(a,h)$. 
Going beyond the
infinite-temperature limit {($\beta=0$)}, we enter {the non-Gibbs regime} within the
system (still for $D \leq 0.5$). That regime can be partitioned to a region
(denoted by green color) where thermalization still takes place but cannot be described
by a Gibbs-type TIO approach. 
Deeper into the nonlinear regime of red-coloring within the $(a,h)$ plane,
our diagnostics developed below suggest a {weakly}  nonergodic, non-Gibbs regime
manifesting persistent localization.
On the other hand, remarkably in the case of $D=2$, there is no region of
TIO applicability and Gibbs-thermalization, for the reasons discussed above.
Here, in the $(a,h)$ plane regions that we have accessed, we have only
been able to identify a {non-Gibbs} thermalization regime
(denoted by the green color).  Finally, in the
red region, once again for $D=2$, 
localization dominates the dynamics and prevents full thermalization on the simulated time scales. 

\begin{figure}[ht]
    \centering
    \includegraphics[width=0.45\textwidth]{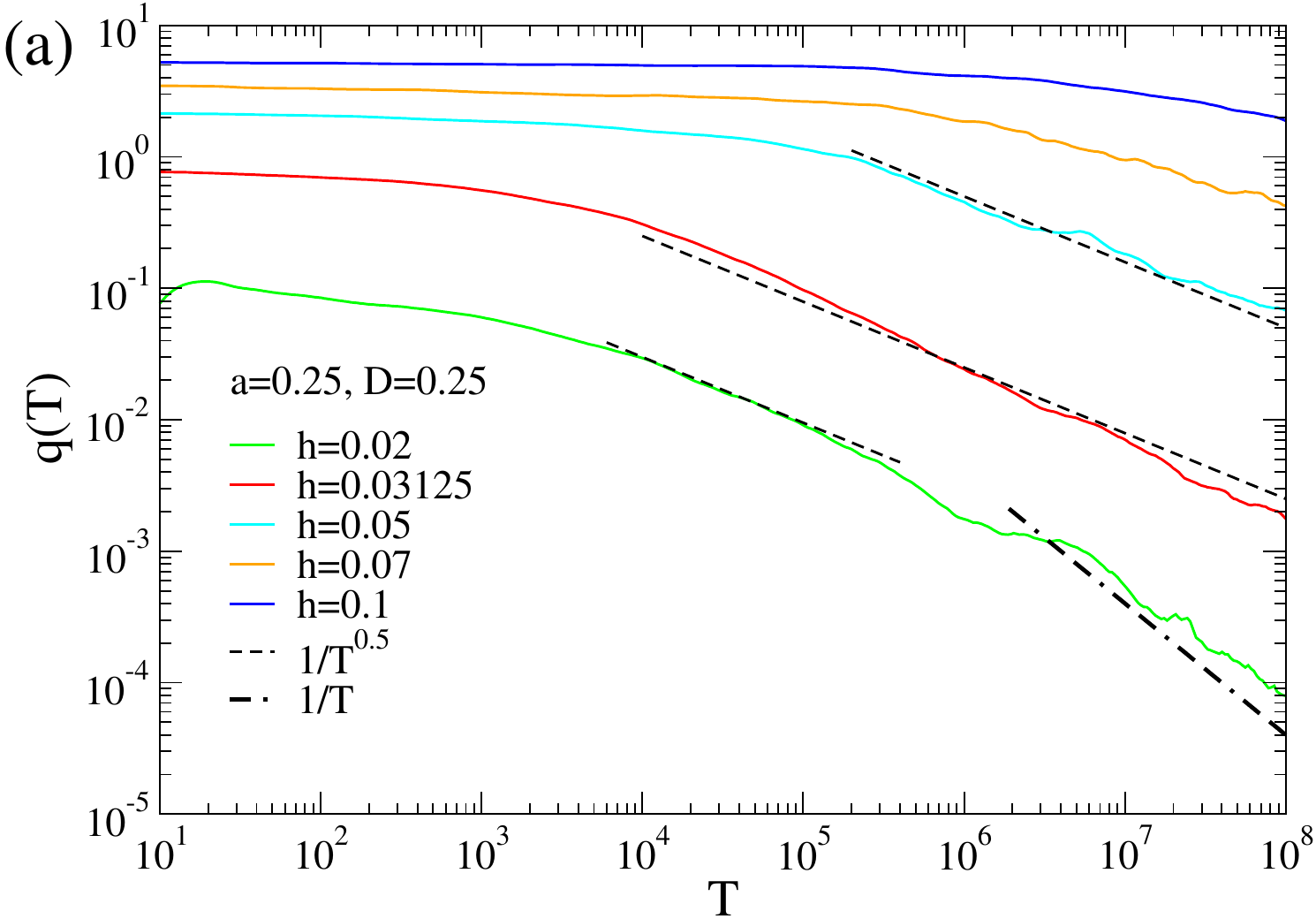}
    \includegraphics[width=0.45\textwidth]{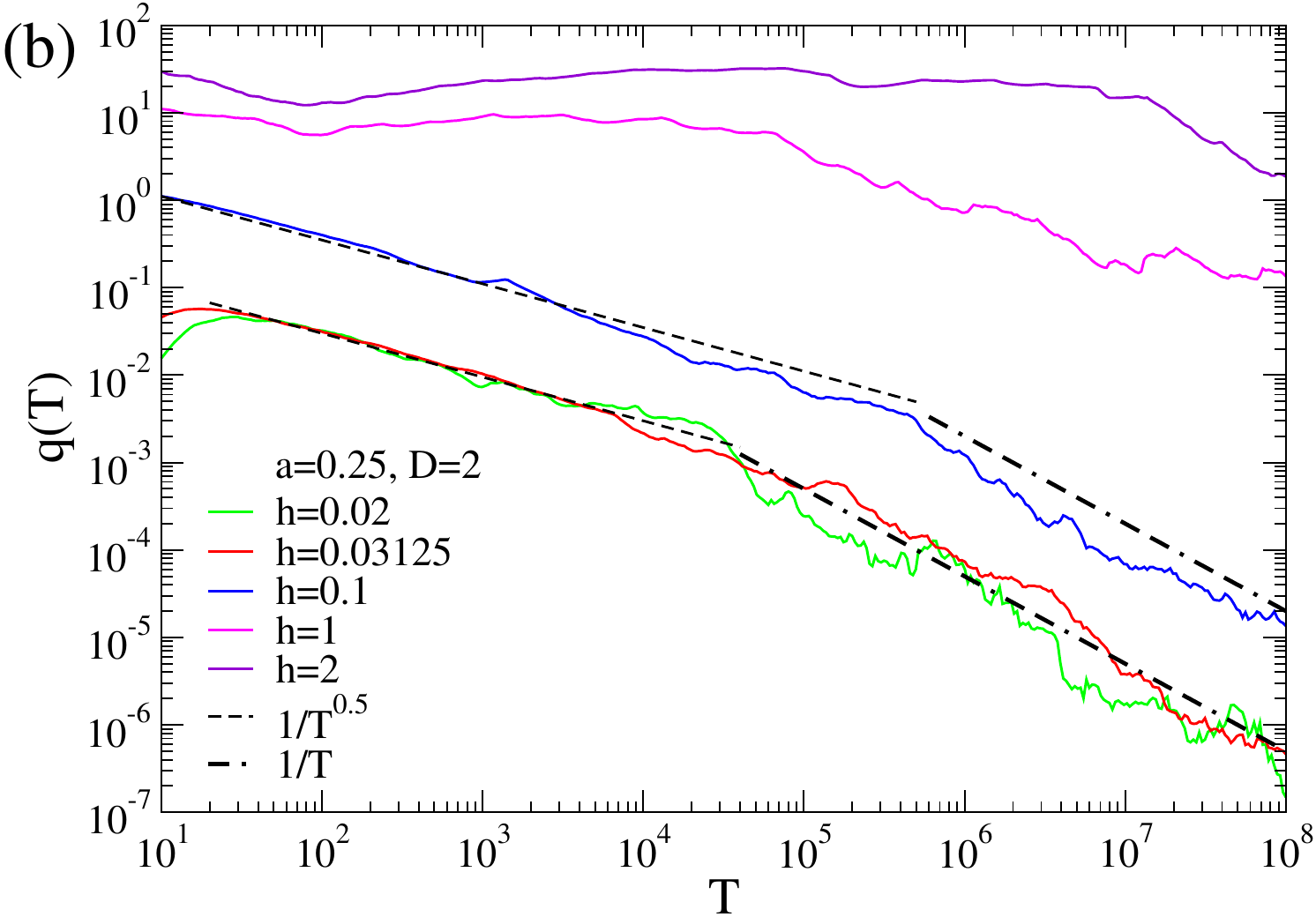} 
\caption{Finite-time averages of the dimensionless variance $q(T)$ for $a=0.25$ with
(a) $D=0.25$ for $h=0.02,\,0.03125,\,0.05,\,0.07,\,0.1$ and
(b) $D=2$ for $h=0.02,\,0.03125,\,0.1,\,1,\,2$. In ergodic regimes, $q(T) \to 0$.
At intermediate times, the decay follows a diffusive scaling
$q(T) \sim T^{-1/2}$,
while at longer times—when accessible—a crossover to a faster
decay $q(T) \sim T^{-1}$ is observed.}
\label{fig:qT}
\end{figure}

\section{Thermalization \& Dynamical Localization: Diagnostics and Results}\label{sec:method}

In this section, we explain the methods that we used to analyze the thermalization of the toy model. Section \ref{sec:finitetimeaverages} describes the  so-called finite-time averages method in detail. This method examines the variance between the ensemble and the time averages of the observables $|b_\ell|^2$, in this
way gauging the potential ergodicity of the system. 
Section \ref{sec:pdf} explains how we study the probability density function of the excursion times of the trajectories of observables off the equilibrium manifold (their ensemble average $a$).
{In section~\ref{sec:pA}, we additionally examine the probability distribution $p(A)$ of local amplitudes $A=|b_\ell|^2$. 
In the Gibbs regime, where the TIO construction is applicable, the analytical prediction is $p(A)=y_0^2(A)$, where $y_0$ is the eigenfunction of the TIO corresponding to the largest eigenvalue. 
This quantity provides a direct test of thermalization by comparing long-time numerical amplitude histograms against the TIO analytical prediction in the Gibbs regime, and reveals the accumulation of large-amplitude excitations in the non-Gibbs and weakly nonergodic regimes~\cite{Rasmussen00,PhysRevE.103.032211}.}

{Finally}, {in section~\ref{sec:compacton},} we study the stability and occurrence of few-site stable localizations in the {weakly} nonergodic regime of different interaction strengths $D$. Assuming a high amplitude few site structure has neighboring sites with amplitudes much smaller than itself, we follow the compacton analysis described in \cite{Kevrekidis23}. We then obtain the pictures of the norm density per site in time $a_\ell(t)$, to confirm if this analytical estimation for compactons is valid and possible for randomized initial states in a large lattice.

In all simulations, we integrated the equations of motion (\ref{eom}) with an explicit adaptive Dormand--Prince Runge--Kutta scheme of order $8(7)$ (\texttt{DOPRI8}, using an embedded error estimator for step-size control). We set $dt=0.1$ as the maximum step size (with the initial trial step size $dt/n_s$); then the integrator adaptively adjusted the internal step size to satisfy the tolerance. We kept the relative norm error ${|A(t)-A(0)|}/{A(0)}$ below $10^{-10}$ and the relative energy error ${|H(t)-H(0)|}/{H(0)}$ below $10^{-8}$, where $A(t)$ and $H(t)$ are the total norm and total energy at time $t$, respectively. 
For our results in this paper, the CPU time to reach $t=10^8$ ranges from ~4 days to ~4 weeks depending on $D$ and the selected densities $a,h$. It decreases as $D$, $a$, and $h$ are reduced.

\subsection{Finite time averages}\label{sec:finitetimeaverages}
We measure the dimensionless variance $q(T)$ over time to evaluate the system’s thermalization by considering the local norm (mass) densities $|b_\ell|^2$ as time-dependent observables. Finite-time averages for $q(T)$ are defined as:
\begin{equation}
  q(T) = \frac{m_2(T)}{m_1^2(T)},  
\end{equation}
where
\begin{align}
    m_1(T) &= \frac{1}{N} \sum_{\ell=1}^N \frac{1}{T} \int_0^T |b_\ell(t)|^2 \, dt, \quad \text{ and } \\
    m_2(T) &= \frac{1}{N} \sum_{\ell=1}^N \left[ \frac{1}{T} \int_0^T |b_\ell(t)|^2 \, dt - m_1 \right]^2.
\end{align}

In thermalized (ergodic) regimes, time and ensemble averages converge and $q(T)$ decays toward zero. For $D<0.5$, where 
{the Gibbs region and its $\beta=0$ boundary are} well-defined, we can probe both Gibbs and non-Gibbs regimes. For $D=0.25$, we find that thermalized cases typically show an initial $\sim 1/\sqrt{T}$ power-law decay before ultimately reaching $\sim 1/T$, as expected for ergodic dynamics~\cite{MithunJJ19,Merab22}. 

Fig.~\ref{fig:qT} illustrates this power-law approach of $q(T)$ to zero in thermalizing systems, with deviations from this trend in some high-energy cases that {subsequently exhibit signatures of weak nonergodicity according to the excursion-time analysis.} 
For $h=0.02$ (green) 
in Fig.~\ref{fig:qT}(a) and {$h=0.02,\,0.03125,\,0.1$} (green, red, blue) in Fig.~\ref{fig:qT}(b), we observe an initial decay $q(T)\sim 1/\sqrt{T}$, followed at longer times by a faster $\sim 1/T$ scaling. For $h=0.03125$ (red) and $h=0.05$ (cyan) in Fig.~\ref{fig:qT}(a), only the $\sim 1/\sqrt{T}$ regime is visible within our time window. 
A $\sim 1/\sqrt{T}$ regime followed by a faster $\sim 1/T$ decay was also reported for a finite-size nonlinear unitary-circuit lattice model~\cite{Merab22}. 
In our simulations, the duration of the intermediate $\sim 1/\sqrt{T}$ regime depends on the initial state and, for fixed $a$, $D$, and $N$, generally increases with the average energy density.
For larger $h$, initial states with $h=0.07,\,0.1$ (orange, blue) in Fig.~\ref{fig:qT}(a) and $h=1,\,2$ (magenta, purple) in Fig.~\ref{fig:qT}(b) exhibit a slower decay of the variance with no clear asymptotic scaling, suggesting strongly delayed ergodization over the investigated time scales.  
To further probe the ergodic character of the dynamics, we study the probability density function of excursion times, and the probability distribution of local amplitudes as described in the next sections \ref{sec:pdf} and \ref{sec:pA}.

\begin{figure}[ht]
    \centering
    \includegraphics[width=0.45\textwidth]{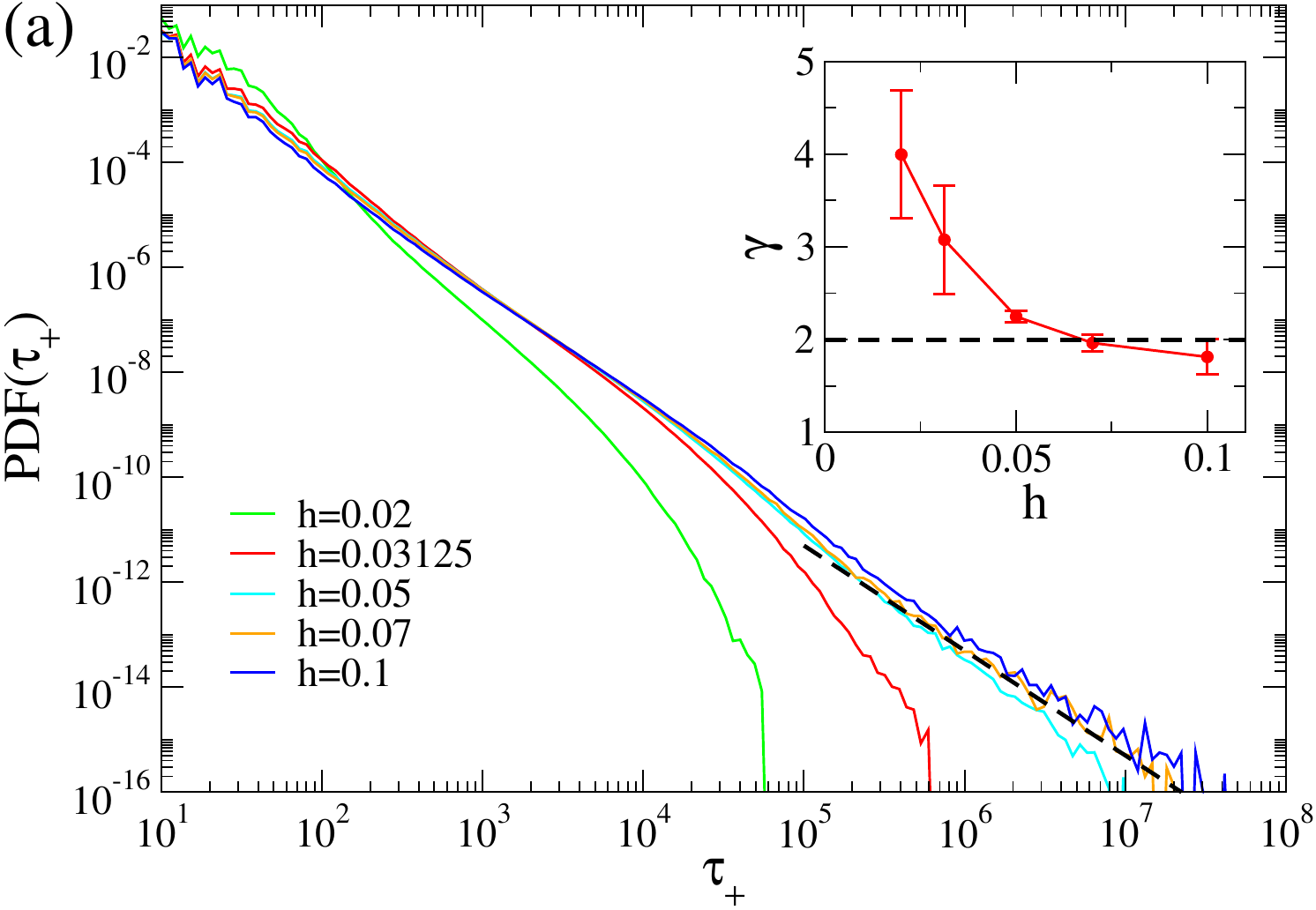} 
        \includegraphics[width=0.45\textwidth]{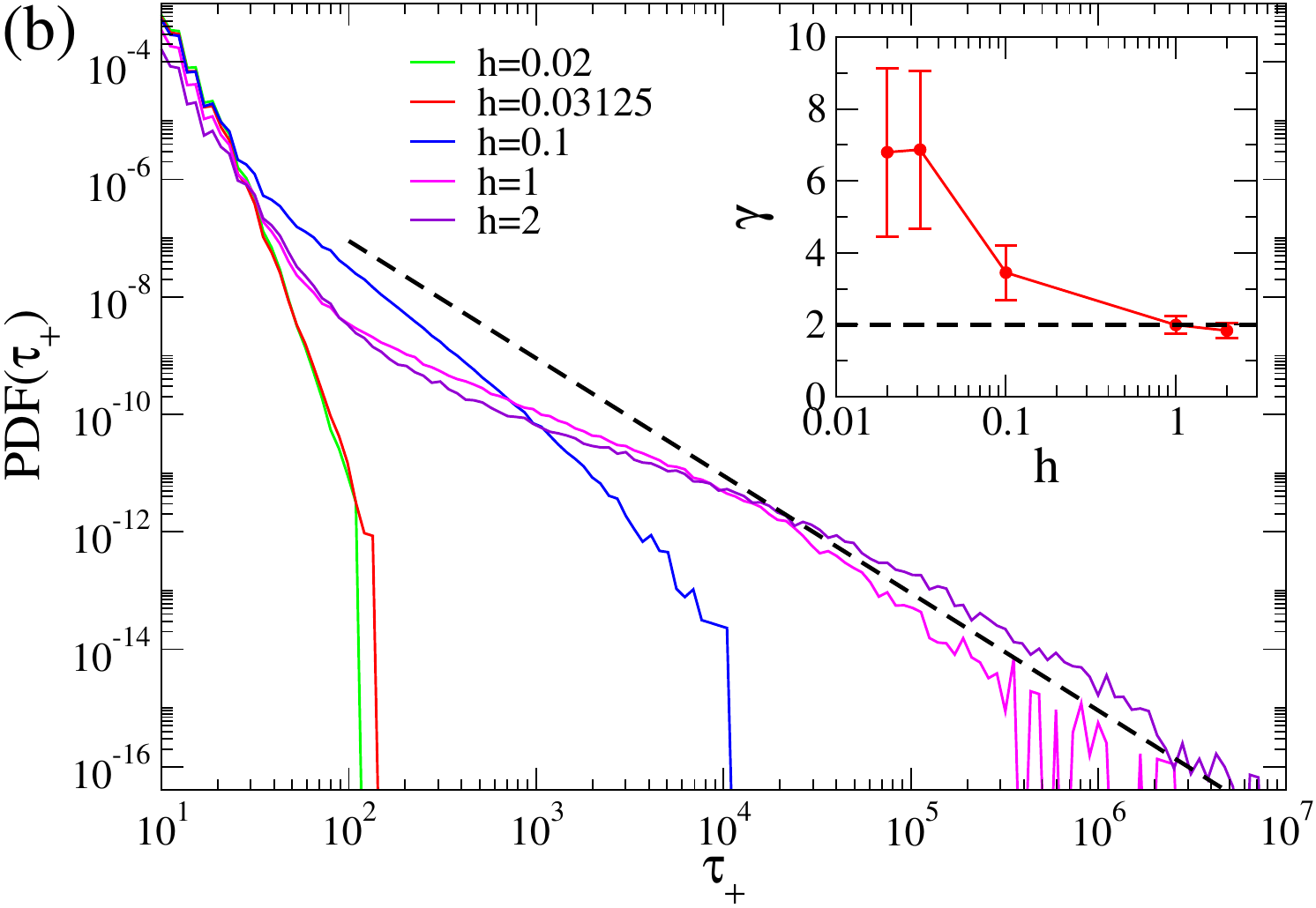}
    \caption{The probability density functions of the excursion times $\tau_+$ 
    for $a=0.25$ with (a) $D=0.25$, and (b) $D=2$ for different energy densities $h$. The black dashed lines in both panels refer to a power law decay with an exponent $\gamma=2$ ($\sim 1/{\tau_+}^2$), which separates the results with exponent $\gamma\leq 2$ (weakly  nonergodic), from those with $\gamma> 2$ (ergodic).  Insets: tail exponent $\gamma(h)$ obtained from the local slope of the log-log PDF. Red circles show the mean of $\gamma(\tau_{+})$ over the last two decades, and error bars its standard deviation.}
    \label{fig:pdfpD}
\end{figure}

\subsection{PDF of excursion times}\label{sec:pdf}
As in the previous section, we choose the local norm densities, $a_\ell = |b_\ell|^2$, as our local observables. The diagnostic developed herein proceeds by calculating the mean norm density $a$ for the entire system and then tracking each $a_\ell(t)$. We count the instances when $a_\ell(t)$ 
crosses the equilibrium value (or so-called the ensemble average of the observables $\langle a_\ell \rangle_\Gamma = a$) and measure the time intervals between these piercings, which are called \textit{excursion times} \cite{Danieli17,MithunDNLS18}. Specifically, we classify the excursion times $\tau$ based on whether $a_\ell > a$ or $a_\ell < a$, denoting them as $\tau_+$ and $\tau_-$, respectively. We then determine the probability distribution function (PDF) for $\tau_+$ and $\tau_-$. 
To compute $\mathrm{PDF}(\tau_{\pm})$, we bin the measured $\tau_{\pm}$ values into 200 bins and normalize by the total number of events and the bin width.

The mean excursion time, in general, can be defined as 
\begin{equation}\label{eq:gamma}
\langle \tau \rangle = \int_{\tau_0}^\infty \tau\, PDF(\tau)\, d\tau \sim \int_{\tau_0}^\infty \tau\, (1/\tau^\gamma)\, d\tau,
\end{equation}
where $PDF(\tau)$ denotes either $PDF(\tau_+)$ or $PDF(\tau_-)$, $\tau_0>0$ is a lower cutoff (a minimal excursion time), and the power-law form is assumed for the tail of $PDF(\tau)$. 
{If the power-law decay exponent $\gamma \leq 2$, the mean of $\tau$ 
diverges; the excursion-time distribution is then heavy-tailed, so that arbitrarily long excursions occur with non-negligible probability, even though every individual excursion remains finite. Thus, we expect $\gamma \leq 2$ for a weakly nonergodic system (see \cite{Danieli17,MithunDNLS18} for details). We use the term \emph{weakly nonergodic} in the sense of weak ergodicity breaking~\cite{Bel2006,Rebenshtok2008,Rebenshtok07,MithunDNLS18}: the dynamics is not strictly nonergodic, which would require the phase space to split into disjoint invariant components protected by additional symmetries, but instead proceeds through a sequence of long-lived metastable states (here, localized excitations) whose lifetimes are broadly distributed with a power-law tail. When the mean lifetime diverges, a trajectory may still eventually explore the accessible phase space, but time averages converge extremely slowly, controlled by rare long-lived excursions. We identify this regime by an excursion-time PDF with tail exponent $\gamma\leq 2$, so that thermalization is not attained within our longest accessible integration times ($t\leq 10^8$).}

While we compute both $PDF(\tau_+)$ and $PDF(\tau_-)$, we focus on 
$PDF(\tau_+)$ to examine the ergodic-to-weakly-nonergodic crossover, as in previous studies \cite{MithunDNLS18,MithunJJ19}. 
This is because we are interested in how much time some trajectories of $a_\ell$ spend at higher densities ($a_\ell(t)>a$), which appear as localizations in time in the norm density dynamics.

For each initial state (see Fig.~\ref{fig:pdfpD}), we extract the tail exponent from the local slope of the log--log representation of $\mathrm{PDF}(\tau_{+})$. 
The local exponent is defined as
\begin{equation}
\gamma(\tau_{+}) \equiv -\,\frac{d\log \mathrm{PDF}(\tau_{+})}{d\log \tau_{+}}.
\end{equation}
To suppress tail fluctuations, the $\log \mathrm{PDF}$ is first smoothed
using a Hodrick–Prescott filter, yielding $\log \widetilde{\mathrm{PDF}}$. The derivative is then estimated by a centered finite difference
along $\log \tau_+$,
\begin{equation}
\gamma_i \approx
-\frac{\log \widetilde{\mathrm{PDF}}_{i+1}
- \log \widetilde{\mathrm{PDF}}_{i-1}}
{\log \tau_{+,i+1} - \log \tau_{+,i-1}}.
\end{equation}
The analysis is restricted to the asymptotic tail region
$\tau_+ \in [10^{-2}\tau_{\max},\, \tau_{\max}]$.
The reported value of $\gamma$ is the mean of
$\gamma(\tau_+)$ over this window,
while the error bar represents its standard deviation,
reflecting residual tail fluctuations rather than fitting uncertainty.

In our simulations, the tails of $PDF(\tau_+)$ for initial states in the Gibbs regime ($\beta>0$) consistently exhibit a power-law decay with exponent $\gamma>2$. This implies finite mean excursion times and supports 
ergodic dynamics in the Gibbs regime. Such behavior is expected within Gibbs statistics 
and has been observed 
for other lattice models, including the DNLS~\cite{MithunDNLS18}, Klein--Gordon~\cite{Danieli19}, and Josephson-junction chains~\cite{MithunJJ19}. An example is shown in Fig.~\ref{fig:pdfpD}(a) [and Fig.~\ref{fig:qT}(a)] for $D=0.25$, $a=0.25$, and $h=0.02$.

More interestingly, we also find $\gamma>2$ for some initial states in the non-Gibbs regime. A representative case ($a=0.25$, $h=0.05$) in the ergodic but non-Gibbs (green) region is shown in Fig.~\ref{fig:qT}(a) and Fig.~\ref{fig:pdfpD}(a). We locate the ergodic--{weakly-}nonergodic crossover by scanning initial states in the $(a,h)$ plane and identifying where the PDF method yields $\gamma=2$. The resulting boundary (solid red line) in Fig.~\ref{fig:phase}(a) does not coincide with the $\beta=0$ line (blue), leaving a significant intermediate region (green) that is ergodic yet non-Gibbs. Similar thermalized but non-Gibbsian regions are known 
in other nonlinear lattice models such as DNLS~\cite{MithunDNLS18} and Salerno \cite{PhysRevE.103.032211}. { Non-Gibbsian regimes in DNLS-like systems have been described in terms of localized high-energy states~\cite{Rumpf2004,Rumpf2009} and microcanonical interpretations of high-energy states~\cite{Buonsante2016BoltzmannGibbsEntropy,Buonsante2017NegativeMicrocanonicalTemperature,MithunDNLS18}. Here, we refer to this region neutrally as the non-Gibbs regime, emphasizing that the grand-canonical Gibbs construction is not normalizable there. }

Fixing the norm density at $a=0.25$ and increasing the energy density $h$ further into the {non-Gibbs} part of the density diagram, we eventually enter the {weakly} nonergodic (red) region in Fig.~\ref{fig:phase}. In this regime, the excursion-time PDF develops a power-law tail with $\gamma\le 2$ [orange and blue curves in Fig.~\ref{fig:pdfpD}(a)], implying diverging mean excursion times. This confirms that {weakly} nonergodic dynamics can occur within the non-Gibbs regime, but it does not cover the entire {non-Gibbs} region.

In Fig.~\ref{fig:pdfpD}(b), we repeat the same initial-state construction as in Fig.~\ref{fig:pdfpD}(a), but now for $D=2$. Using the same $(a,h)$ values as in Fig.~\ref{fig:pdfpD}(a) (i.e., $0.02\le h\le 0.1$, including the cases that are {weakly} nonergodic for $D=0.25$), we find $\gamma>2$ for all these runs, indicating ergodic dynamics at $D=2$. 
This suggests that the stronger nearest-neighbor interaction at larger $D$ enhances chaos and promotes thermalization. To access weakly nonergodic behavior at $D=2$, we therefore increase the energy density further at fixed $a$, which leads to $\gamma\le 2$ for $h=1$ and $h=2$ [Fig.~\ref{fig:pdfpD}(b)], consistent with the weakly nonergodic (red) region in the $(a,h)$ phase diagram of Fig.~1(b). 
In these high-energy cases, the system does not thermalize 
within our longest integration times, $t\le 10^8$.

\begin{figure}[ht]
    \centering
        \includegraphics[width=0.45\textwidth]{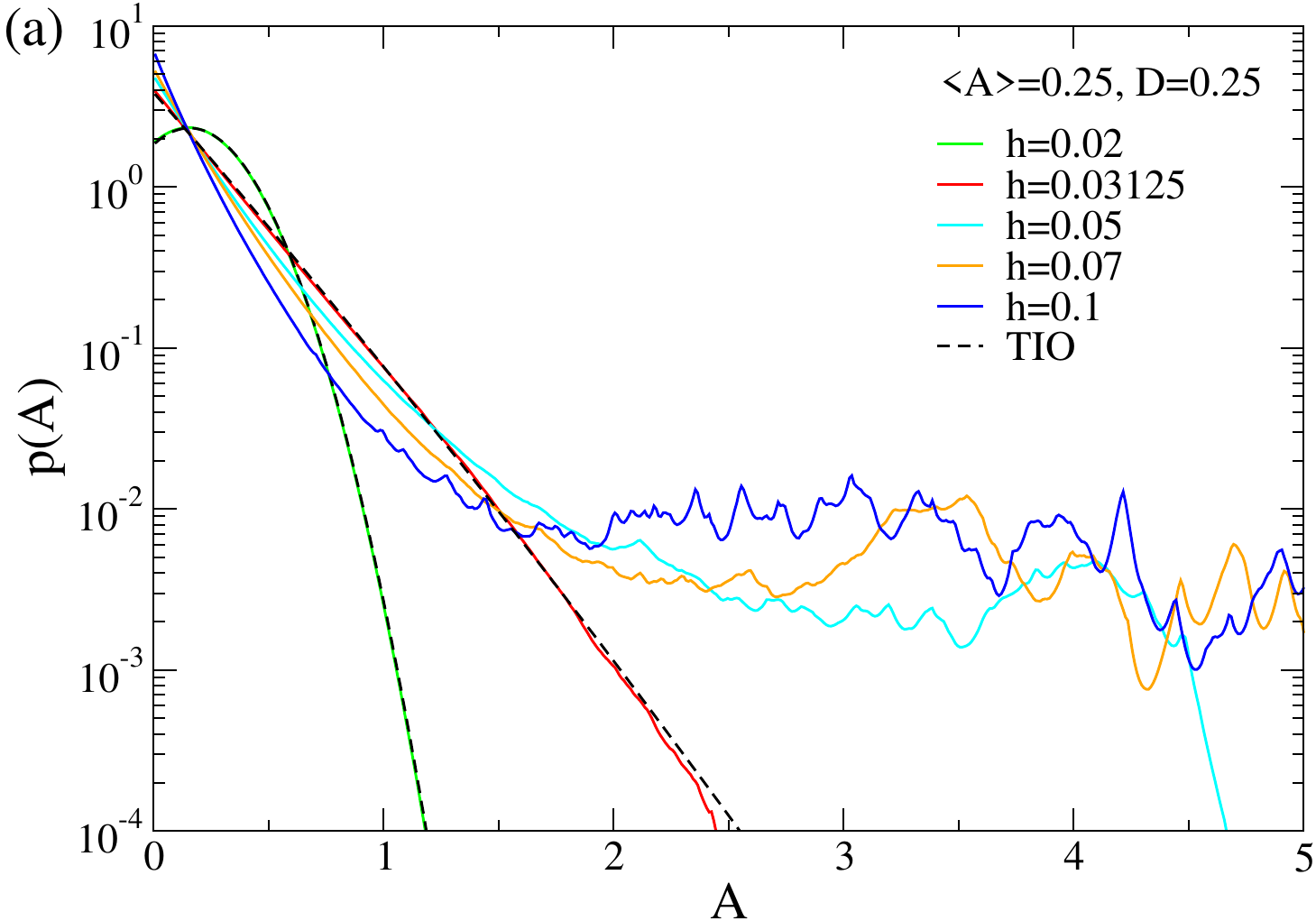} 
        \includegraphics[width=0.45\textwidth]{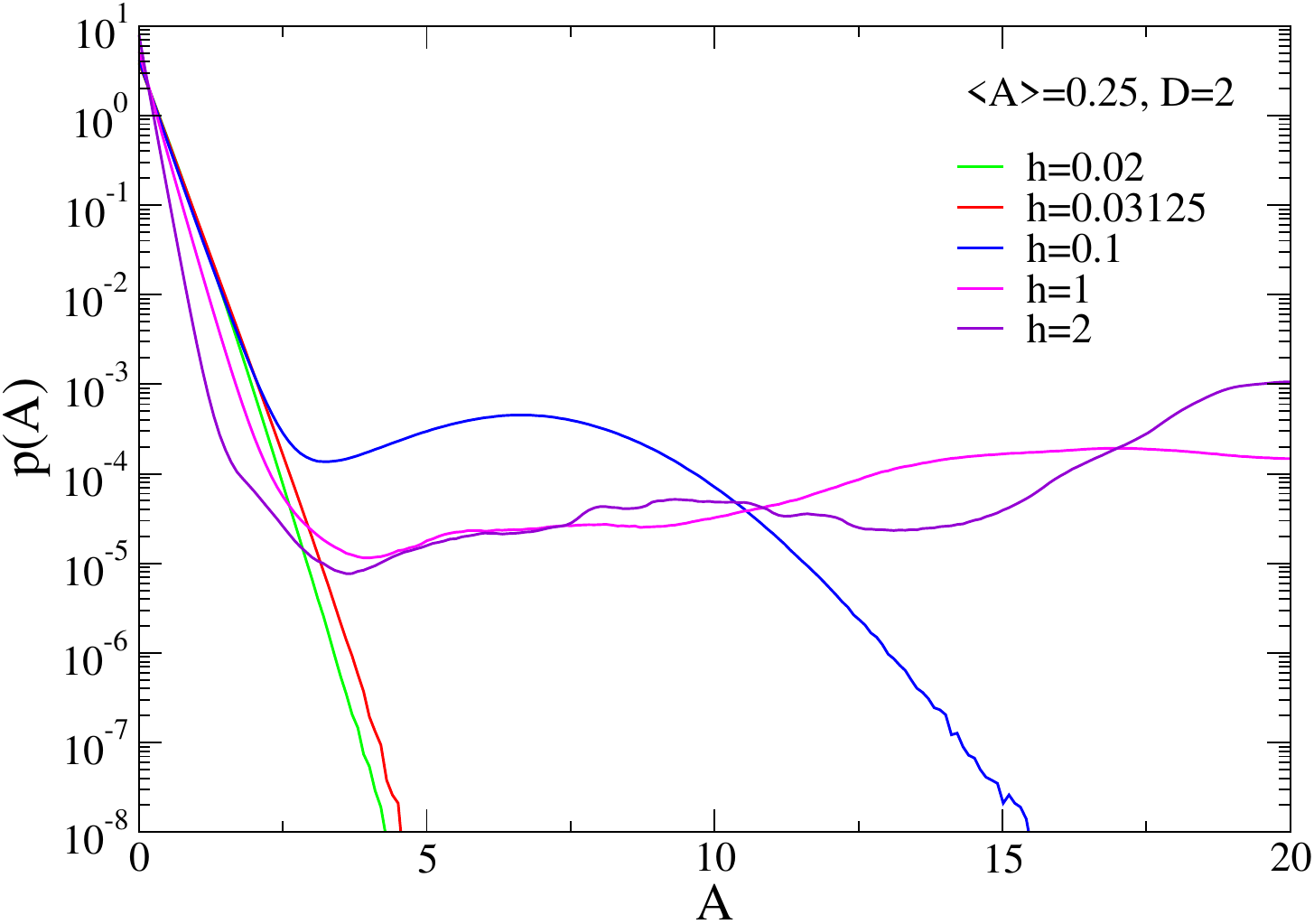}        
    \caption{Probability distribution $p(A)$ of the local amplitudes $A=|b_\ell|^2$ for different energy densities $h$ at fixed average norm density $a=0.25$: (a) $D=0.25$; (b) $D=2$. The numerical curves are obtained from histograms accumulated over all sites and over the full simulation time, up to $t=10^8$. In panel (a), the transfer-integral-operator (TIO) results for $h=0.02$ and $h=0.03125$ agree with the numerical distributions, confirming the Gibbs-equilibrium character of the numerically obtained states.}   
    \label{fig:pA_A}
\end{figure}

\begin{figure}[ht]
  \centering
  \includegraphics[width=.35\textwidth]{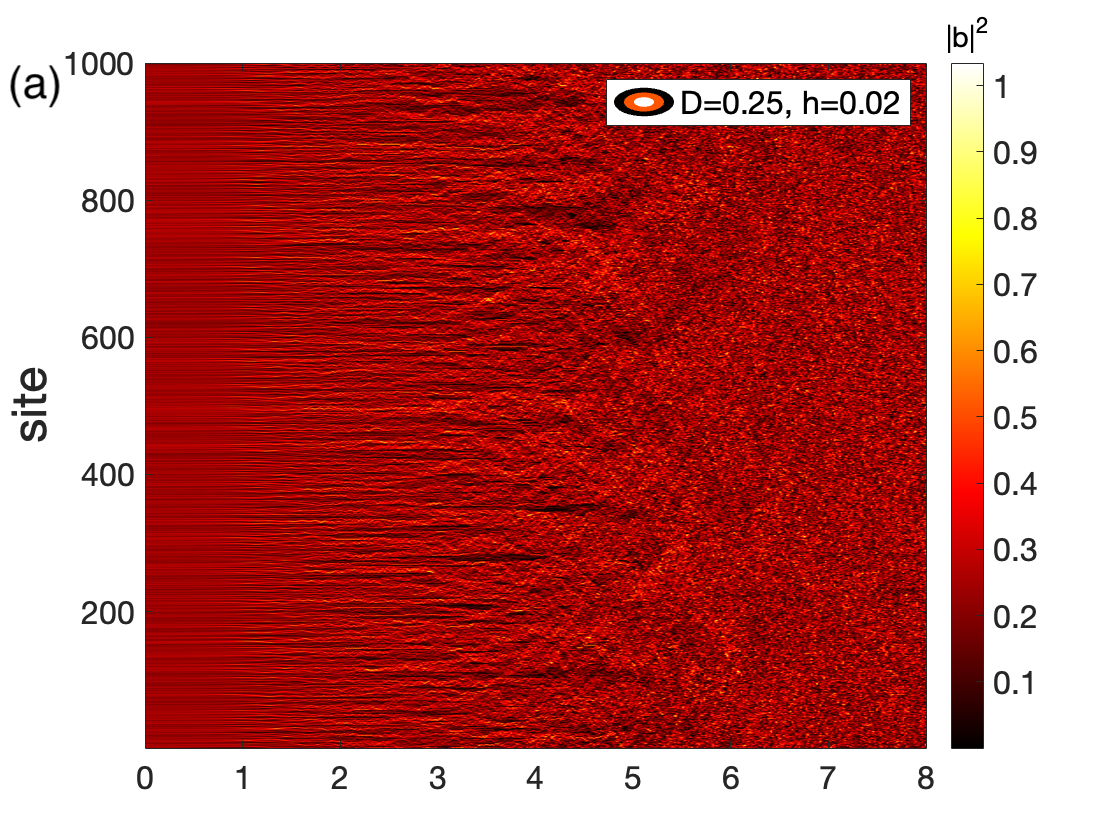}\\[-9.2pt] 
  \includegraphics[width=.35\textwidth]{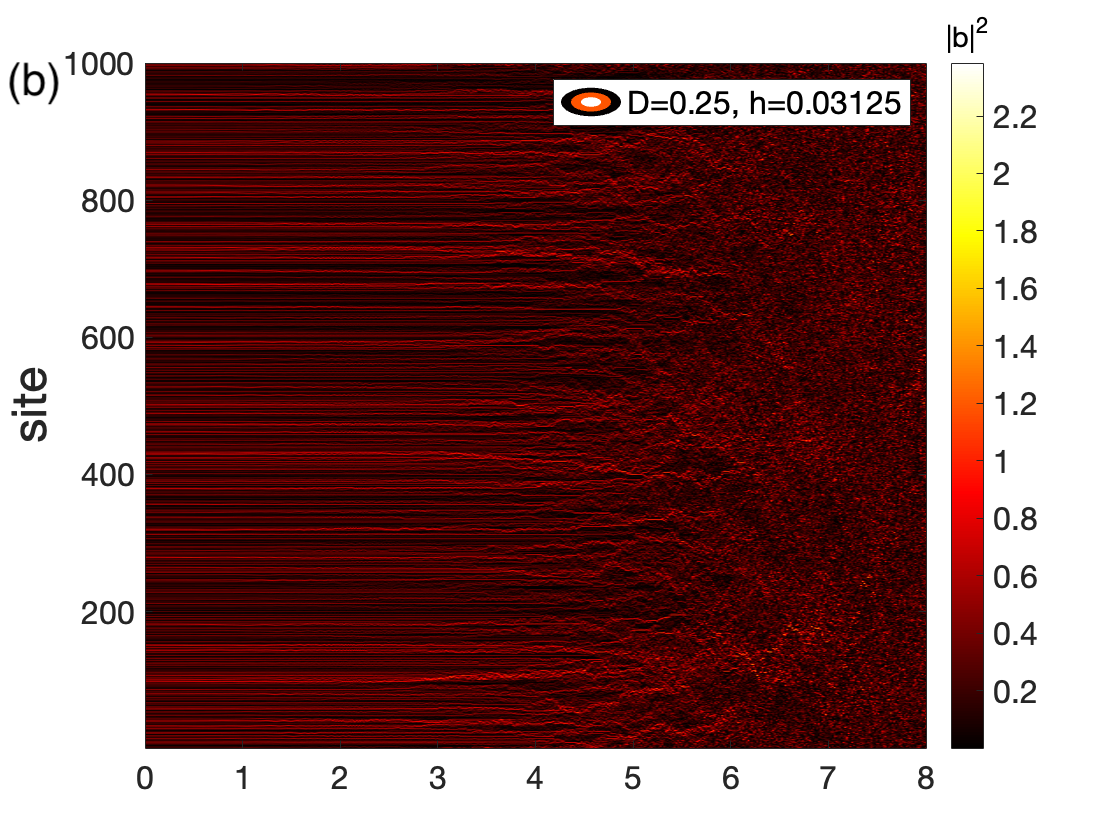}\\[-9.25pt] 
  \includegraphics[width=.35\textwidth]{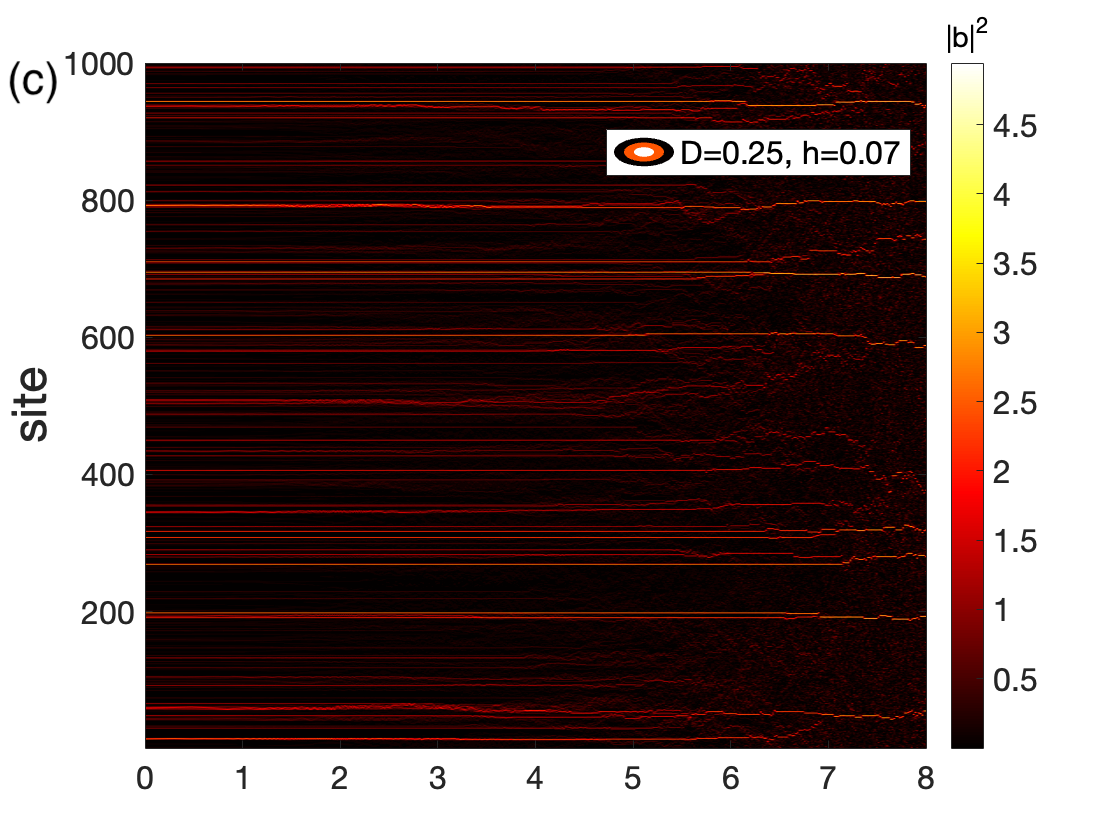}\\[-9.25pt]
  \includegraphics[width=.35\textwidth]{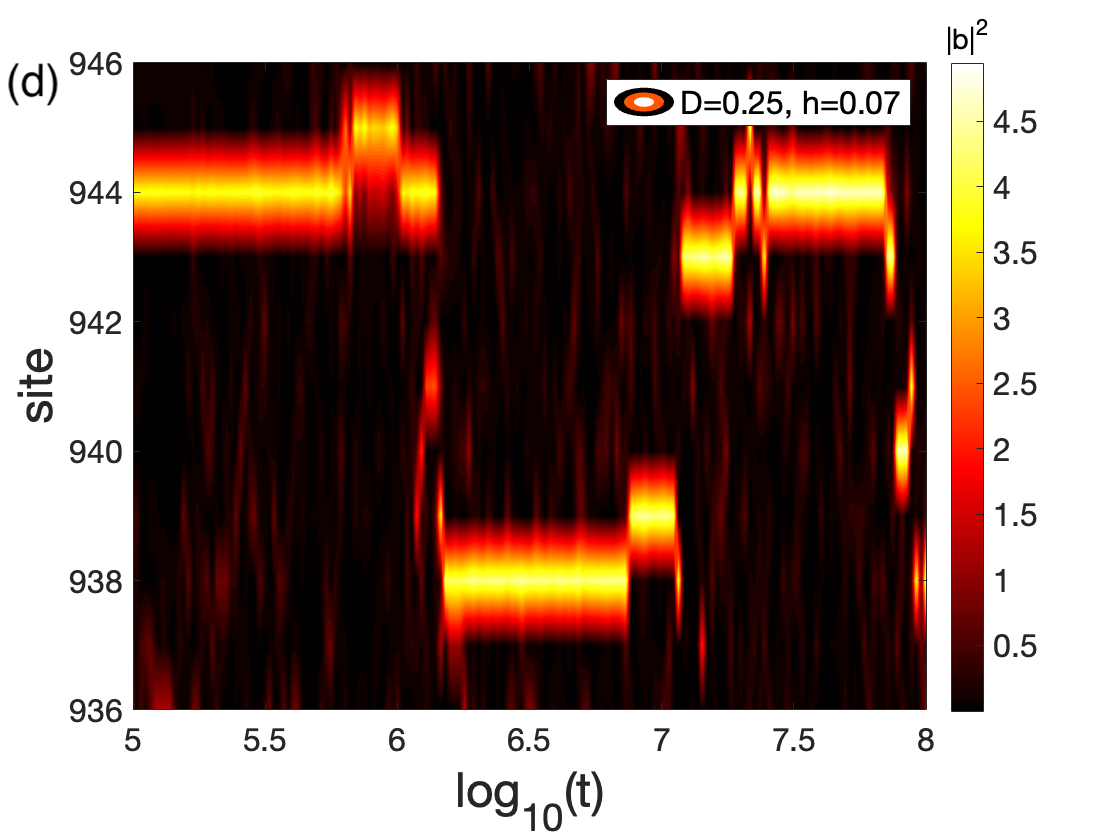} 
  \vspace{-0.4cm}
\caption{Increasing energy density reduces chaotic mixing and leads to long-lived localization at fixed $D=0.25$ and $a=0.25$. Local norm densities $a_\ell(t)$ are color coded in time for (a) $h=0.02$ ($\beta>0$), (b) $h=0.03125$ ($\beta=0$), and (c) $h=0.07$ {(non-Gibbs)}. Part (d) shows a zoomed view of (c), where one-site localization persists for long times.}
\label{fig:snap_D025a025h002}
\end{figure}

\subsection{Distribution of local amplitudes}\label{sec:pA}

{To further characterize the thermalization properties, we examine the probability density function $p(A)$ of the local amplitudes $A=|b_\ell|^2$. 
Numerically, $p(A)$ is obtained from long-time histograms of the site amplitudes. In the Gibbs regime, it can also be computed analytically within the transfer-integral-operator (TIO) framework~\cite{Rasmussen00,PhysRevE.103.032211}, where
\[
p(A)=y_0^2(A), \qquad \int_0^\infty p(A)\,dA = 1,
\]
and $y_0$ is the normalized eigenfunction associated with the largest eigenvalue $\lambda_0$ of the transfer-integral kernel. 
Accordingly, the agreement between the numerical histogram and the TIO prediction provides a direct test of relaxation towards the corresponding Gibbs equilibrium state. Beyond the Gibbs regime, where no such TIO equilibrium reference is available, the numerical distribution $p(A)$ remains a useful marker of localization. A heavier tail or an enhanced secondary peak at large $A$ indicates an increasing contribution from high-amplitude localized excitations, which is a characteristic of weakly nonergodic regimes~\cite{Rasmussen00,PhysRevE.103.032211}.

The results are shown in Fig.~\ref{fig:pA_A}. For $D=0.25$,  $a=0.25$, the two lowest energy cases ($h=0.02$ and $h=0.03125$) fall within the Gibbs regime and show very good agreement between the numerical histograms and the TIO prediction [Fig.~\ref{fig:pA_A}(a)], confirming relaxation to Gibbs equilibrium. 
As $h$ increases in the non-Gibbs regime, the distribution 
develops a progressively enhanced weight at large $A$, reflecting the growing role of persistent high-amplitude excitations. 
This trend is already visible for $h=0.05$, which lies close to the 
ergodic--{weakly} nonergodic crossover identified by 
$\mathrm{PDF}(\tau_+)$ and is still classified as ergodic by that 
diagnostic, yet $p(A)$ reveals the onset of localized structures. 
For $h=0.07$ and $0.1$, this weight grows further, consistent with 
the {weakly} nonergodic character signaled by $q(T)$ and 
$\mathrm{PDF}(\tau_+)$ in sections~\ref{sec:finitetimeaverages} and~\ref{sec:pdf}.

In Fig.~\ref{fig:pA_A}(b), for $D=2$, we do not have an analytical reference distribution because the transfer-integral kernel diverges for $D>0.5$ (see section~\ref{sec:model}). 
Nevertheless, the numerically obtained $p(A)$ remains informative. At low energies ($h=0.02,\,0.03125$), the distribution decays rapidly, consistent with ergodic mixing. For $h=0.1$, the distribution develops a small secondary peak before eventually decaying, yet the overall decay to negligible values indicates that this case remains ergodic. At high energies ($h=1,\,2$), the distribution develops a pronounced secondary structure and persistent weight at large $A$, in direct correspondence with the {weakly} nonergodic regime identified by $q(T)$ and $\mathrm{PDF}(\tau_+)$.}  Thus, the local-norm-density distributions provide a complementary way to distinguish thermally distributed fluctuations from regimes dominated by persistent compact localization.

\subsection{Mechanism for few-site localization}\label{sec:compacton}

For the {weakly} nonergodic regime, long-time integrations of Eqs.~\eqref{eom}--\eqref{hamiltoniantoy} show that the norm density $a_\ell=|b_\ell|^2$ becomes strongly localized: it concentrates on a single site for $D<1$, and, typically, on two  adjacent sites for $D>1$. We verified this trend for $0.01\le D\le 2$ (not shown).

\begin{figure}[ht]
\centering
\includegraphics[width=0.4\textwidth]{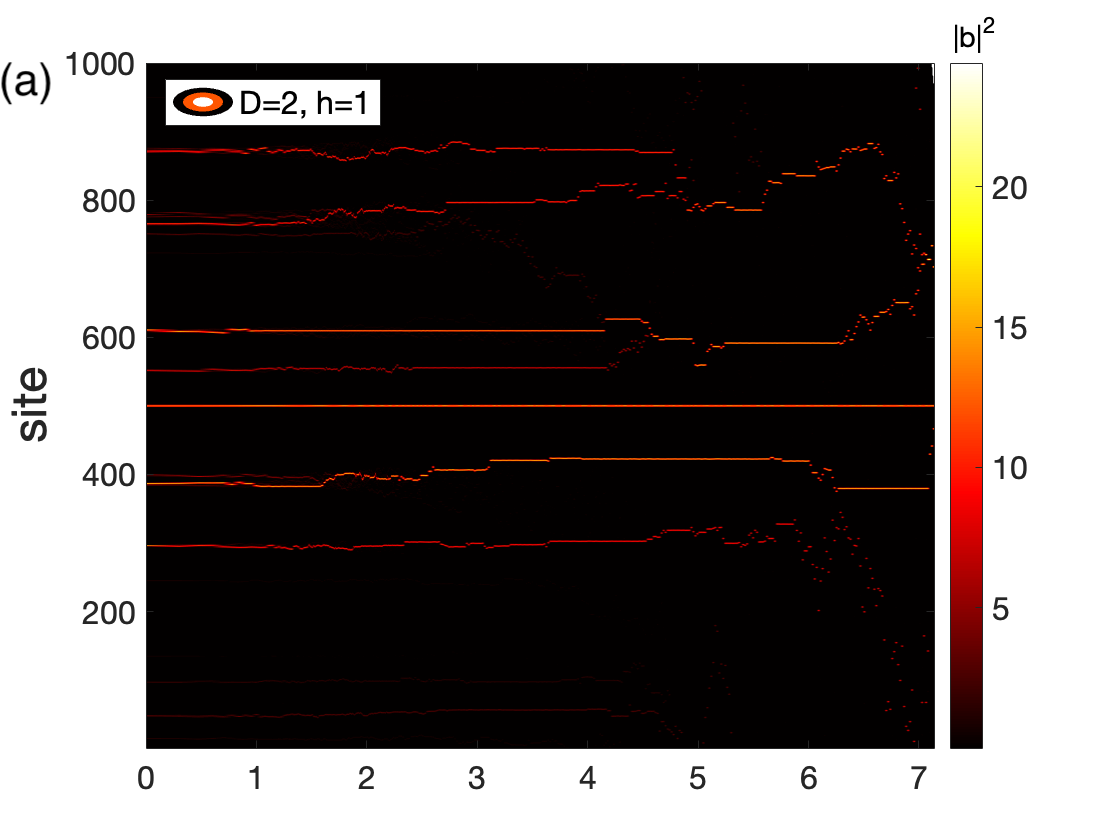}\\ [-9.25pt] 
\includegraphics[width=0.4\textwidth]{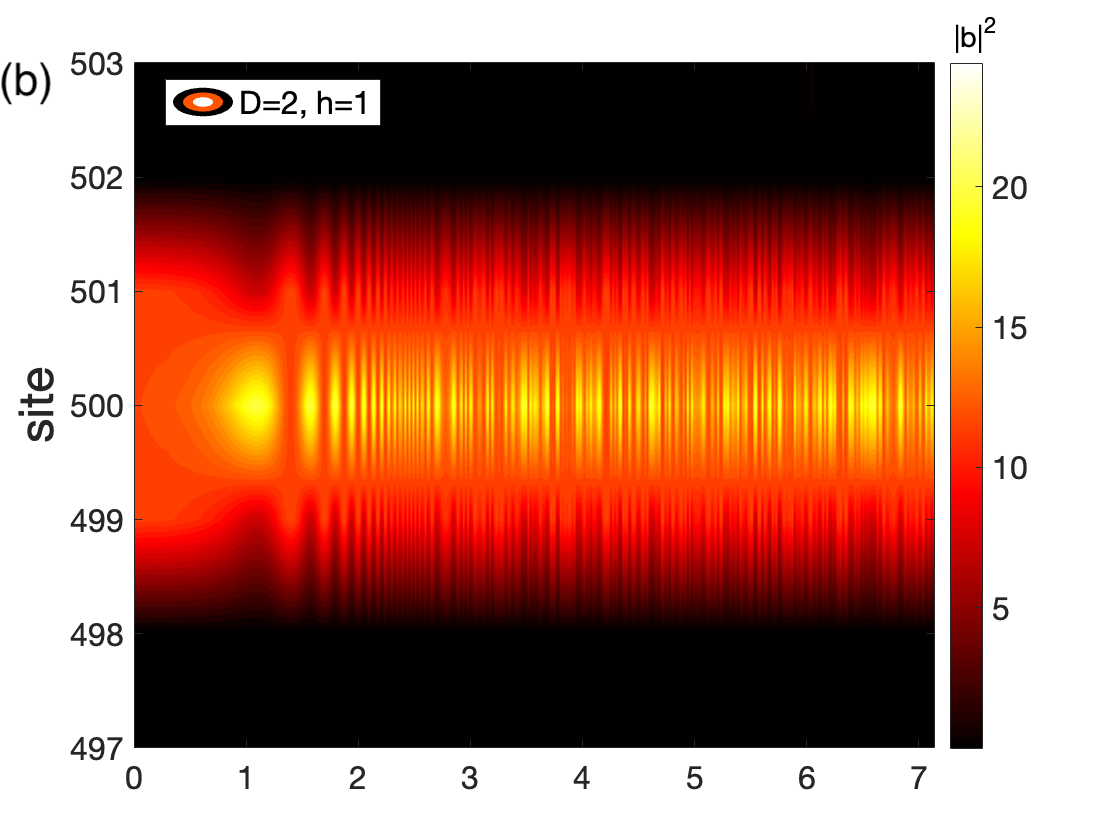} \\[-9.25pt] 
\includegraphics[width=0.4\textwidth]{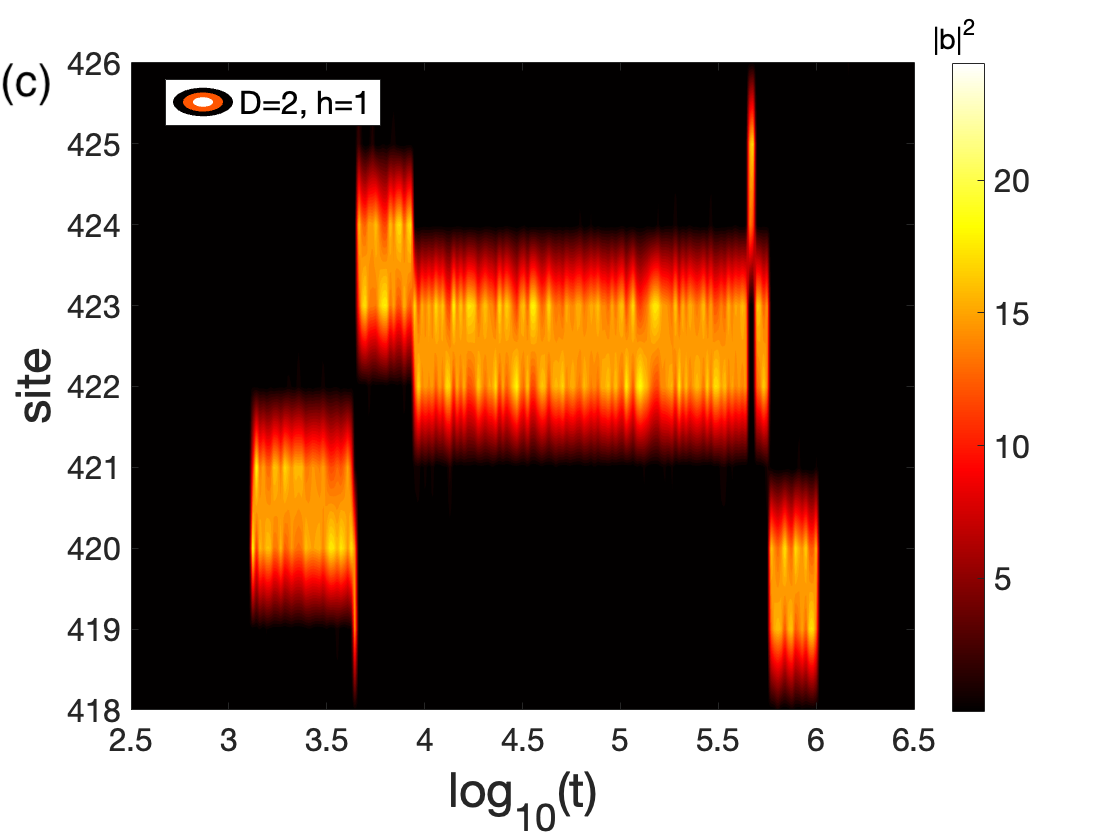}  \\[-4pt] 
\caption{High-energy localization on two and three sites for $D=2$. Local norm densities $a_\ell$ are color-coded in time for $a=0.25, h=1$. (a) Full lattice. (b) Zoom of (a) showing a 3-site, in-phase localization. (c) Zoom of (a) showing the dominant 2-site staggered localization. The initial state is prepared according to Eq.~(\ref{eq:compact}) with $a=0.25$, $x=19$, and $c=21$.}
    \label{fig:snapD2a025h1zoom}
\end{figure}

\begin{figure}[ht]
    \centering  
    \includegraphics[width=0.45\textwidth]{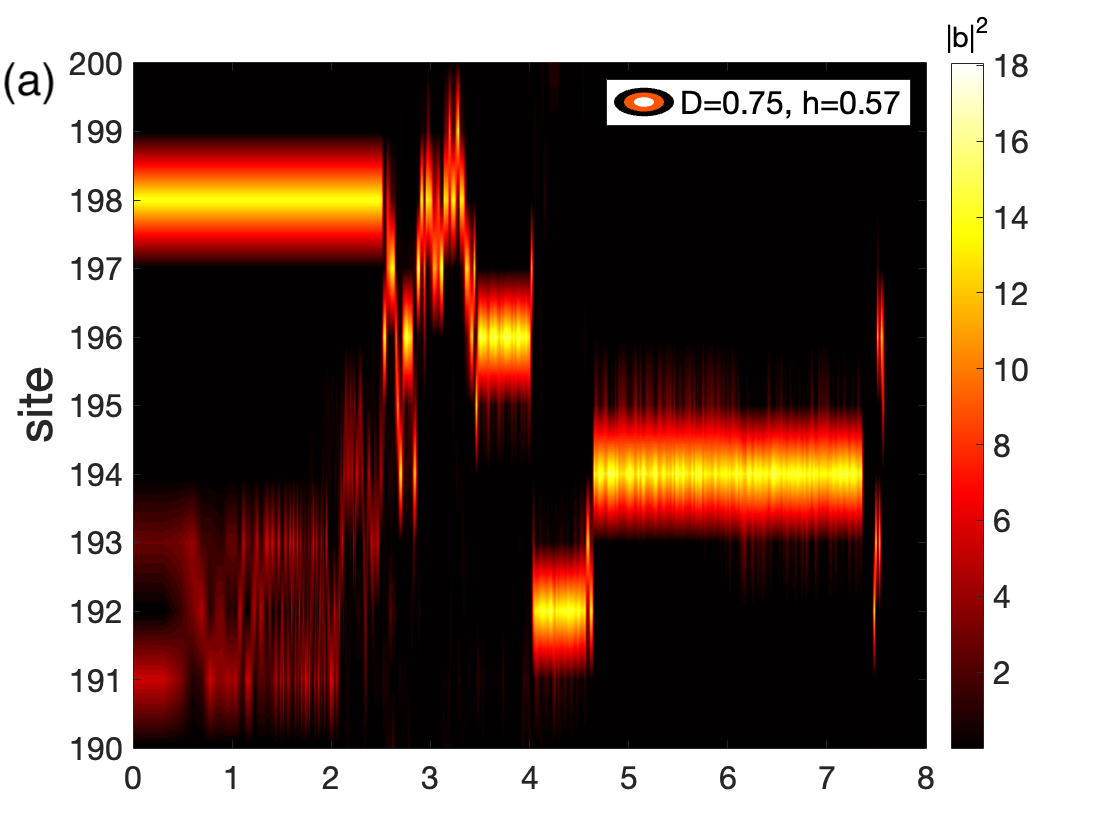} \\[-9.25pt] 
    \includegraphics[width=0.45\textwidth]{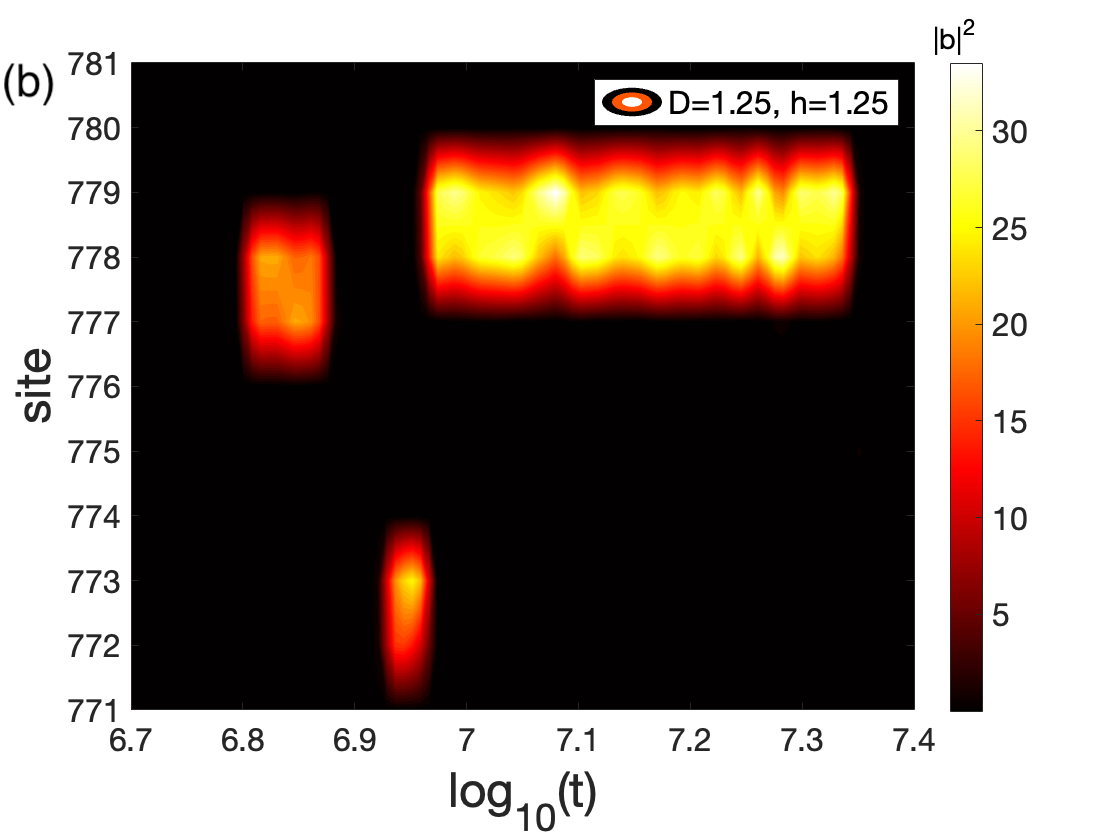}
\caption{Weakly nonergodic dynamics exhibits a one-site compacton for $D<1$ and a two-site staggered compacton for $D>1$. Local norm densities are shown as a function of time for $a=0.25$: (a) $D=0.75$, $h=0.57$ and (b) $D=1.25$, $h=1.25$. The initial states are prepared using Eq.~(\ref{eq:ic2}) with the same random realization of $\eta_\ell$ and $\epsilon_\ell$, using $(c,x)=(4,12)$ in (a) and $(c,x)=(10,18.13)$ in (b). We plot only a subset of lattice sites (a zoom in $\ell$) to highlight the localization pattern.}
\label{fig:snapD0.75D1.25}
\end{figure}

A natural explanation comes from the standing-wave compactons supported on $m$ sites ($m=1,2,3$) analyzed in \cite{Kevrekidis23,Colliander13}. These solutions assume that all sites outside the compact support are exactly zero, whereas our initial data excite the entire lattice. Nevertheless, at a high energy for fixed total norm (an initial state with a high $H/A^2$), the dynamics {\it spontaneously} concentrates 
a significant portion of the total norm on the block(s) of one or few adjacent sites, while the rest of the chain carries small-amplitude oscillations. These localized structures can drift along the chain over the course of the evolution [see, e.g., Figs.~\ref{fig:snap_D025a025h002}(d), \ref{fig:snapD2a025h1zoom}(c), and \ref{fig:snapD0.75D1.25}(a,b)], although the extent of this mobility depends on the parameter regime. In line also with the argumentation of~\cite{MarzuolaMattingly2025}, 
in the {Gibbs regime ($\beta>0$)}, one would expect this to  single out 
{\it energy minimizers}, while in the {non-Gibbs regime} to localize
the mass into {\it energy maximizers}. 
Accordingly, in the {weakly} nonergodic regime, the compacton energies provide accurate local energetic guidance towards the localized profiles that should be observed in 
the long-time evolution of the system.

We therefore proceed to evaluate the Hamiltonian \eqref{hamiltoniantoy} on standing-wave compactons supported on $m$ sites with compact support 
$S=\{\ell_0,\ldots,\ell_0+m-1\}$ for $m=1,2,3$, and $b_\ell=0$ for $\ell\notin S$. On $S$ we define 
\[
b_\ell=\sqrt{A_\ell}\,e^{i\phi_\ell},\qquad A_\ell\ge 0,\qquad \sum_{\ell\in S}A_\ell=A.
\]
Only adjacent sites within $S$ enter the interaction term in \eqref{hamiltoniantoy}, for each adjacent pair $(\ell,\ell-1)$
\[
\mathrm{Re}\!\big[(\bar b_\ell)^{2} b_{\ell-1}^{2}\big]=A_\ell A_{\ell-1}\cos\!\big(2(\phi_\ell-\phi_{\ell-1})\big).
\]
Following \cite{Kevrekidis23}, we call $\phi_\ell-\phi_{\ell-1}=0$ a real "in-phase" compacton and $\phi_\ell-\phi_{\ell-1}=\pi/2$ a "staggered" compacton. In this paper, we will focus more on the existence of staggered compactons on a lattice, as we investigate the {weakly} nonergodic initial states with a high $H/A^2$. The work of~\cite{Kevrekidis23}
examined energy minimizers earlier (see, e.g., their Fig.~7) (see also the discussion
of~\cite{MarzuolaMattingly2025}); however, in our {weakly nonergodic, non-Gibbs} regime,
it is the maximizers (which are often suitable staggered configurations) that become
central to the dynamics.

\smallskip
\noindent\textbf{1 site.} 
$S=\{\ell_0\}$, $A_{\ell_0}=A$. There are no adjacent sites within the support, so the interaction term in \eqref{hamiltoniantoy} vanishes:
\begin{equation}
H_{1}=\frac{1}{4}\,A^2\qquad(\text{independent of }D).
\label{eq:H1_final}
\end{equation}

\smallskip
\noindent\textbf{2 sites.} 
$S=\{\ell_0,\ell_0+1\}$, set $A_{\ell_0}=x$, $A_{\ell_0+1}=y$ with $x+y=A$. 
Then
\[
H(x,y)=\frac{1}{4}(x^2+y^2)\mp\frac{D}{2}xy,
\]
where the upper ($-$) sign corresponds to the in-phase (real) compacton ($\Delta\phi=0, \cos{(2\Delta\phi)}=1$) and the lower ($+$) sign to the staggered compacton ($\Delta\phi=\pi/2, \cos{(2\Delta\phi)}=-1$). Extremizing under $x+y=A$ (by symmetry or a Lagrange multiplier) gives $x=y=A/2$, hence
\begin{equation}
H_{2,\mathrm{in}}(D)=\frac{1-D}{8}\,A^2,\qquad
H_{2,\mathrm{stag}}(D)=\frac{1+D}{8}\,A^2.
\label{eq:H2_final}
\end{equation}

\smallskip
\noindent\textbf{3 sites.}
$S=\{\ell_0-1,\ell_0,\ell_0+1\}$, set $A_{\ell_0-1}=A_{\ell_0+1}=x$, $A_{\ell_0}=y$, with $2x+y=A$.
The reduced energy is
\[
H(x,y)=\frac{1}{2}x^2+\frac{1}{4}y^2 - D\,x y \cos(2\Delta\phi).
\]
For the staggered compacton ($\Delta\phi=\pi/2$), a {Lagrange-multiplier extremization under the constraint $2x+y=A$ yields $x=(1-D)A/(3-4D)$ and $y=(1-2D)A/(3-4D)$ (for $D\neq 3/4$), and the corresponding energy}
\begin{equation}
H_{3,\mathrm{stag}}(D)=\frac{1-2D^2}{4(3-4D)}\,A^2.
\label{eq:H3_final}
\end{equation}

For the in-phase configuration ($\Delta\phi=0$), the analogous three-site calculation gives $x=\frac{1+D}{4D+3}\,A$, $y=\frac{1+2D}{4D+3}\,A$, and
\begin{equation}\label{eq:H3in}
H_{3,\mathrm{in}}(D)=\frac{1-2D^2}{4(4D+3)}\,A^2.
\end{equation}

\begin{figure}[ht]
    \centering
    \includegraphics[width=0.45\textwidth]{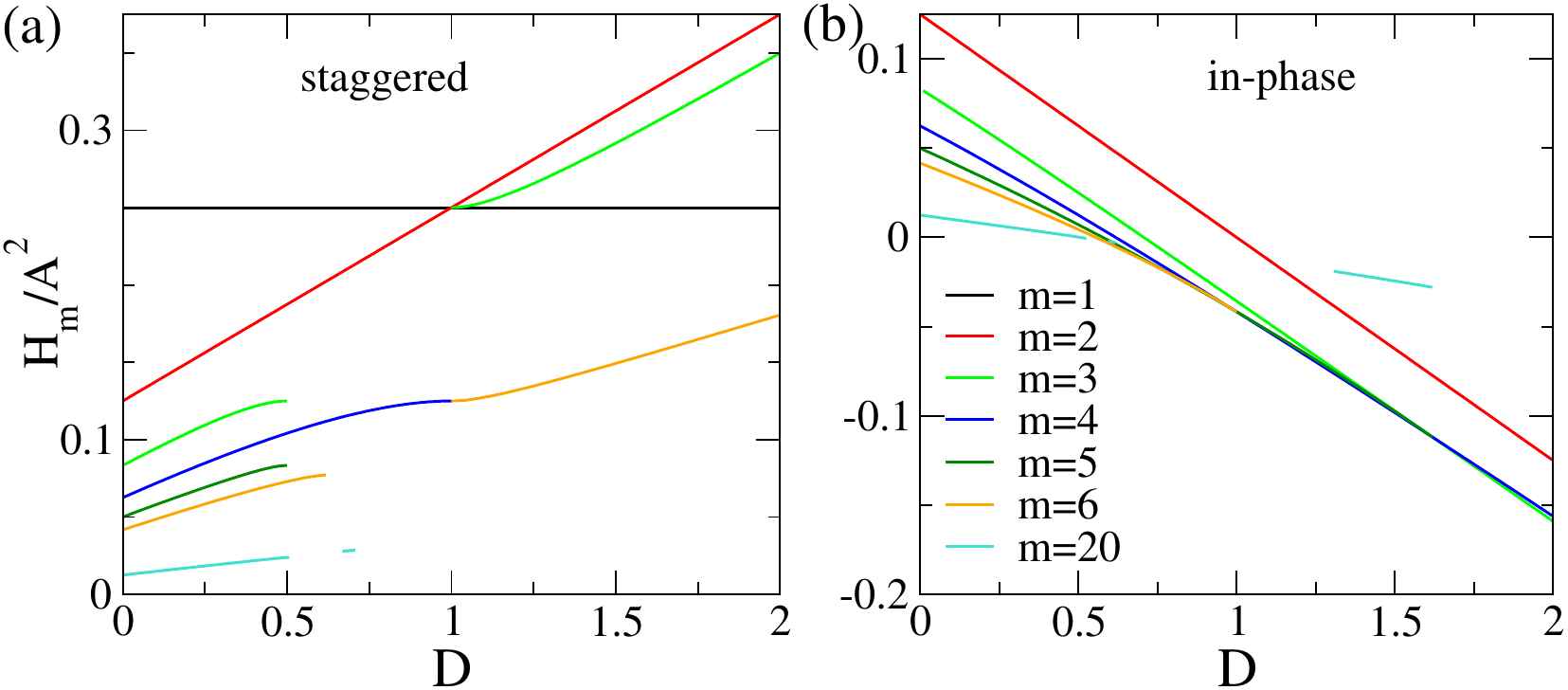} 
    \includegraphics[width=0.45\textwidth]{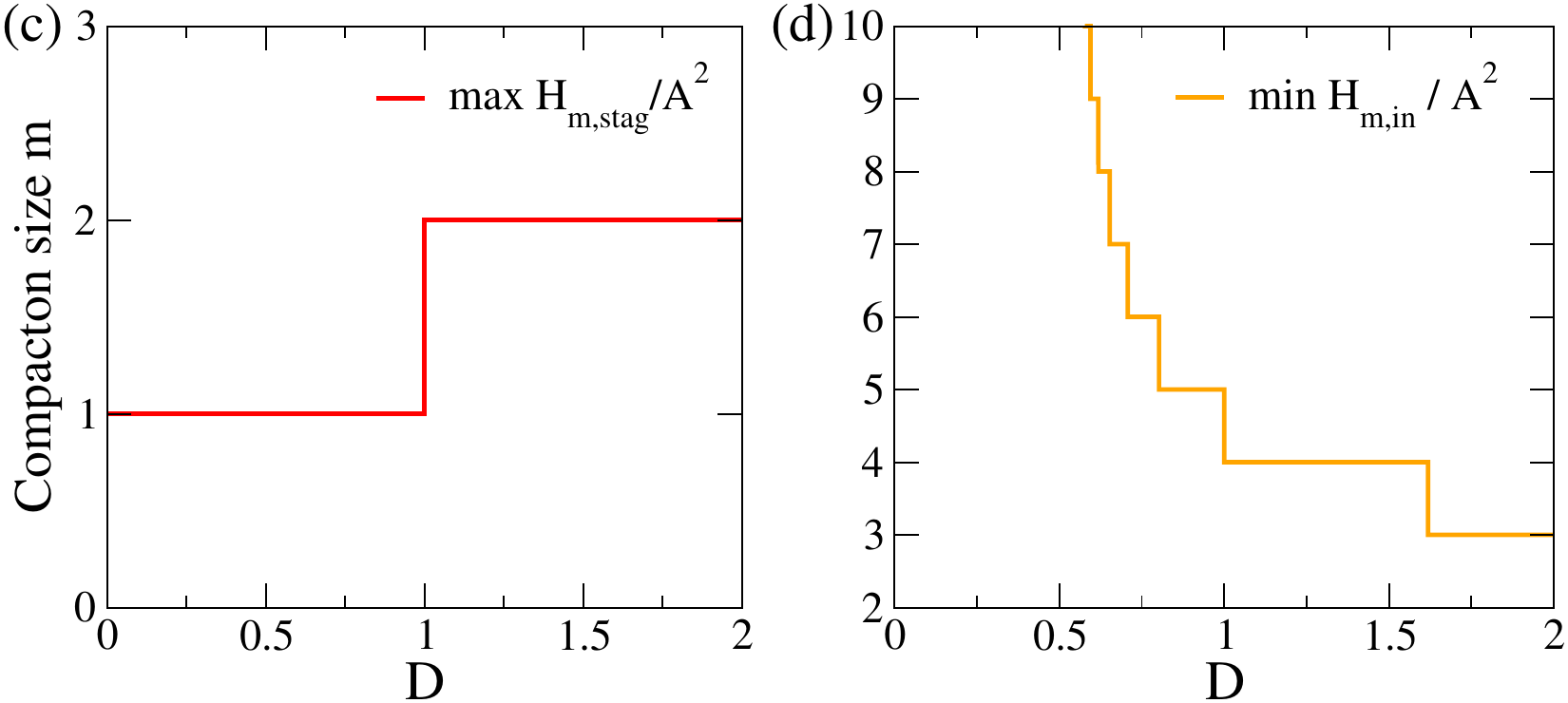}
\caption{Reduced energies $H_m(D)/A^2$ at fixed total norm $A$ from Lagrangian extremization. Panels (a, b): representative staggered [$\Delta\phi=\pi/2$] and in-phase [$\Delta\phi=0$] branches for $m=1,\dots,6,20$. Panels (c, d): $\arg\max_m H_{m,\mathrm{stag}}(D)$ and $\arg\min_m H_{m,\mathrm{in}}(D)$ from a full scan over $m=1,\dots,100$. Intervals with negative stationary amplitudes or a singular reduced system are omitted.}
\label{fig:energymaximizer}
\end{figure}

From Eqs.~(\ref{eq:H1_final})--(\ref{eq:H3in}), we obtain the energy ordering at fixed $A$:
\begin{align}
    &D < 1:\quad 
    H_{3,\mathrm{in}} < H_{2,\mathrm{in}} < H_{3,\mathrm{stag}} < H_{2,\mathrm{stag}} < H_{1}, 
    \label{eq:dless1} \\[4pt] 
    &D > 1:\quad 
    H_{3,\mathrm{in}} < H_{2,\mathrm{in}} < H_{1} < H_{3,\mathrm{stag}} < H_{2,\mathrm{stag}} .
    \label{eq:dgreater1}
\end{align}
{Equations~(\ref{eq:dless1})--(\ref{eq:dgreater1}) thus identify the constrained energy maximizer at fixed $A$ as the one-site state for $D<1$ and as the two-site staggered state for $D>1$, with $H_{3,\mathrm{stag}}$ only marginally below $H_{2,\mathrm{stag}}$ in the latter case [see Fig.~\ref{fig:energymaximizer}(a)]. At the marginal point $D=1$ the one-, two-, and three-site staggered branches are degenerate, $H_{1}=H_{2,\mathrm{stag}}=H_{3,\mathrm{stag}}=A^2/4$, as also visible in Fig.~\ref{fig:energymaximizer}(a).

In the weakly nonergodic regime the long-time dynamics indeed selects these constrained maximizers, confirming the analytically determined energy-maximizing support $m$ obtained over $0\le D\le 2$ in Fig.~\ref{fig:energymaximizer}(c). Figures~\ref{fig:snap_D025a025h002}, \ref{fig:snapD2a025h1zoom}, and \ref{fig:snapD0.75D1.25} show $a_\ell(t)$ snapshots displaying one-site localization for $D=0.25$ and $D=0.75$, and a two-site staggered profile for $D=1.25$ and $D=2$; the intermediate $D=0.75$ and $D=1.25$ initial states were prepared with Eq.~(\ref{eq:ic2}) at parameters that place them in the weakly nonergodic regime (Fig.~\ref{fig:snapD0.75D1.25}). The one-site outcome for $D<1$ parallels findings in the DNLS and Josephson-junction chains~\cite{MithunDNLS18,MithunJJ19}. The degeneracy at $D=1$ does not translate into degenerate dynamics: the generic high-$H/A^2$ evolution still produces one-site localization, while two- and three-site states emerge only when the initial data already excite $m=2$ or $m=3$ adjacent sites with appropriate phases through Eq.~(\ref{eq:compact}).}

{
To test whether compactons with larger support can compete with the few-site configurations discussed above, we extended the same Lagrangian extremization to $m=1,\ldots,100$ (derivations and code can be found in~\cite{kati_data_allv}). 

Figures~\ref{fig:energymaximizer}(a,b) show the resulting energies $H_m(D)/A^2$, while Figs.~\ref{fig:energymaximizer}(c,d) show the corresponding optimizing compacton size $m$. 
Note that some curves in Fig.~\ref{fig:energymaximizer} have gaps for certain $D$ intervals. These correspond to cases where at least one compacton amplitude $A_\ell$ becomes negative or $H_m(D)$ has a zero denominator. 
Figures~\ref{fig:energymaximizer}(a,c) show that the energy-maximizing staggered compacton has size $m=1$ for $D<1$ and $m=2$ for $D>1$; no larger support with $m\leq100$ yields a higher energy. The in-phase family, shown in Fig.~\ref{fig:energymaximizer}(b,d)] has a richer structure: the optimal size is $m=5$ just below $D=1$, $m=4$ for $1\leq D\lesssim1.618$, and $m=3$ for $D\gtrsim1.618$, {consistent with the rigorous result of~\cite{MarzuolaMattingly2025} for $D=2$}. As $D\to1/2^{+}$, the minimizing support grows rapidly and already 
exceeds $m=100$ for $D\lesssim0.5009$, {consistent with the conjecture of~\cite{Kevrekidis23} that no finite-support in-phase minimizer exists for $D\leq1/2$.}

In the deep non-Gibbs regime at high energies (e.g., see the red regions for $D=0.25$ and $D=2$ in Fig.\ref{fig:phase}), our simulations show spontaneous formation of the staggered maximizers (one- and two-site compactons) in the $D$ ranges predicted in Fig.~\ref{fig:energymaximizer}(c). From numerous simulation results, we show representative cases in Figs. \ref{fig:snapD0.75D1.25} and \ref{fig:snap_D025a025h002}, in agreement with the analytical result. Since we did not observe in-phase minimizers of Fig.~\ref{fig:energymaximizer} spontaneously arising from generic initial conditions, we instead tested their stability, by preparing the initial state with Eq.~(\ref{eq:compact}).
The size of the energy-minimizing compacton varies for different $D$ ranges as shown in Fig.~\ref{fig:energymaximizer}(d) (see also \cite{Kevrekidis23}). 
We focus on $D=2$, because the three-site minimizer has been mathematically established for this case~\cite{MarzuolaMattingly2025}.  
We have tested initial states with $m=3,\ldots,20$, and observed that only $m=3$ persists over the long-time evolution (see Fig.~\ref{fig:snapD2a025h1zoom}(b)). This is consistent with its being the in-phase energy minimizer at $D=2$ as shown in Fig.~\ref{fig:energymaximizer}(d), and with the rigorous result of Ref.~\cite{MarzuolaMattingly2025}. }

\section{Discussion}\label{sec:discussion}

In summary, the toy model demonstrates thermalization across a broad parameter space, but not all thermalized regions conform to the traditional Gibbs statistical framework. As illustrated in {green in} Fig.~\ref{fig:phase}, we observe a parameter region where ergodicity is achieved, although Gibbs statistics fails to describe it accurately. {This suggests that a more complete statistical description is required. 
For $D=0.25$, this ergodic regime lies above the $\beta \rightarrow 0$ line, which corresponds to the non-Gibbs part of the parameter space.  
Deep in the non-Gibbs regime, we identify} an ergodic-to-{weakly}-nonergodic {crossover} by examining probability density functions of excursion times $\mathrm{PDF}(\tau_+)$, finite-time averages $q(T)$, {the probability distribution of local amplitudes $p(A)$}, and long-time norm localizations.

As the nonlinear intersite interaction $D$ increases, {we observe a substantial change in the dynamics for the same $(a,h)$ parameter regions.   
In Fig.~\ref{fig:phase}, note that the horizontal axes differ in range and that the ergodic region widens considerably (for the range of amplitudes shown in panel (b)) 
from (a) $D=0.25$ to (b) $D=2$.  
Figures~\ref{fig:qT}, ~\ref{fig:pdfpD}, ~\ref{fig:pA_A} further compare the two cases regarding their thermalization with different methods.} 

In Fig.~\ref{fig:qT}, we observe that the fluctuations of dimensionless variance $q(T)$ are visually more pronounced at larger $D$. 
In the ergodic regimes, the long-time decay crosses over from an initial $\sim T^{-1/2}$ to a faster $\sim T^{-1}$ trend, and this crossover occurs at earlier times for larger $D$. 
This indicates that stronger nonlinear coupling promotes a faster approach to the $T^{-1}$ scaling associated with ergodic dynamics. Our simulations also suggest that increasing $D$ expands the region of $(a,h)$ space where ergodic, non-Gibbs behavior is observed.

{Fig.~\ref{fig:pdfpD} shows the PDF of excursion times $\tau_+$ for varying energy densities $h$. At sufficiently large $h$, the tails are well described by a power law, $P(\tau_+)\sim \tau_+^{-\gamma}$ with $\gamma$ shown as a function of $h$ in the inset. When $\gamma\leq2$, the corresponding mean excursion time $\langle \tau_+\rangle$ diverges in the asymptotic power-law regime, indicating 
heavy-tailed dynamics with rare but statistically important long excursions. As $h$ is decreased, $\gamma$ increases, the power-law regime narrows, and the distributions decay more rapidly at large $\tau_+$ (red and green curves). The heavy tails observed at large $h$ are consistent with long-lived localized structures, such as breathers and compactons, which coexist with mixing in the remaining lattice and strongly delay thermalization. 
Together with the finite-time variance $q(T)$, the local norm-density distribution $p(A)$, and the evolution of $a_\ell(t)$, the excursion-time PDF enables
a clear identification of the ergodic-to-weakly-nonergodic crossover shown in
Fig.~\ref{fig:phase}, which the other diagnostics considered here do not resolve as sharply.}

{
The probability distribution of local amplitudes $p(A)$, with $A=|b_\ell|^2$,
shown in Fig.~\ref{fig:pA_A}, complements this picture. In the Gibbs regime ($D=0.25$, $h\leq 0.03125$), the numerical $p(A)$ agrees very well with the TIO prediction $p(A)=y_0^2(A)$, confirming thermal equilibration. As $h$ increases into the non-Gibbs and weakly nonergodic regimes, the tail of $p(A)$ at large $A$ becomes progressively heavier, reflecting the growing contribution of persistent high-amplitude localized excitations. For $D=2$, where no TIO reference exists, the same qualitative trend is observed: a rapidly decaying $p(A)$ at low energies and a heavier-tailed form at high energies ($h=1,\,2$) associated with long-lived compact localization, in direct correspondence with the weakly nonergodic regime identified by $q(T)$ and $\mathrm{PDF}(\tau_+)$. Taken together, the three diagnostics paint a coherent and mutually reinforcing picture: $q(T)$ characterizes relaxation toward ergodic behavior, $\mathrm{PDF}(\tau_+)$ supplies the operational criterion for weak nonergodicity, and $p(A)$ directly reveals the accompanying accumulation of persistent large-amplitude localized excitations.
}

To observe how the norm is distributed on the lattice in time,
as we progress from an ergodic to a {weakly} nonergodic system, in Fig.~\ref{fig:snap_D025a025h002}, we increase the energy density $h$ of the lattice from (a) 0.02 ($\beta>0$) to (b) 0.03125 ($\beta=0$), and to (c) $0.07$ {(non-Gibbs)}, while the norm density $a=0.25$ and coupling $D=0.25$ are kept fixed. 
In Fig.~\ref{fig:snap_D025a025h002}(d), we have a closer look to the one-site localization in the case of (c) $h=0.07$. 
Although it {appears}  
as a yellow dot near $T=10^8$ on site 938, this is actually a long-time localization on site 938 {from} $T=10^{7.9}$ to $T=10^{8}$, which corresponds to a time interval of nearly $2\times  10^7$. It is accordingly evident
that single-site localization dominates the (very) long system evolution dynamics, in 
line with our discussion of the previous section.

As $D$ increases above $0.5$, the Gibbsian thermalization region altogether
disappears (in line with the divergence of the TIO). In the latter regime, 
we observe non-Gibbsian thermalization{, which we infer} from alternative diagnostics such as finite-time averages and excursion time PDFs. In addition to this regime
of non-Gibbsian thermalization, for sufficiently strong nonlinearities
(large $H/A^2$), we notice that the system favors persistent localization, 
{concentrated on energy maximizers} 
that switch from single
site localization for $D<1$, to two-site (indeed staggered) localization
for $D>1$.

For $D = 0.25$, the system shows predominantly 1-site localization (see Fig.~\ref{fig:snap_D025a025h002}), whereas for $D = 2$, we observe 2-site and 3-site stable localization (see Fig.~\ref{fig:snapD2a025h1zoom}). 
The 2-site (staggered) structures appear spontaneously and are generically favored, although, as shown in Fig.~\ref{fig:snapD2a025h1zoom}, an initialization of a 3-site {\it in-phase} excitation yields a
structure that is quite long-lived, suggesting connections with the work of~\cite{MarzuolaMattingly2025} on energy minimizers. 
These findings are consistent with previous work \cite{Kevrekidis23}, which identified multi-site compactons in similar nonlinear systems. The behavior of the toy model for different values of $D$ correlates well with the theoretical predictions of localization patterns
based on energy maximizers.

\section{Conclusions \& Future Challenges}
\label{sec:conclusion}

We investigated the statistical properties of a toy model previously derived from the continuum {two-dimensional} defocusing NLS model as a simplified representation of
potential turbulent cascades in the latter. In this effective lattice model, the nodes are not actual physical lattice sites but rather groups of wavenumbers.
We identify both ergodic and weakly nonergodic regimes through numerical methods and, where relevant, support this picture analytically through the compacton-energy analysis and
quasi-analytically via the transfer-integral-operator approach. 
Our results highlight a triplet of possible {dynamical} regimes when $D<0.5$, which gives way to two distinguishable regimes when $D>0.5$.

The most elementary regime, consistent with earlier studies~\cite{Rasmussen00,kev09,MithunDNLS18,BALDOVIN20211}, is that of Gibbsian thermalization. The TIO method relates the conserved energy and norm densities to their thermodynamic counterparts, mapping $(a,h)$ to temperature and chemical potential.  
For $D=0.25$, beyond the $\beta=0$ boundary of the standard Gibbs description, we identify a non-Gibbs region in which the dynamics may nevertheless remain ergodic. The finite-time variance of the local norm density exhibits a characteristic $1/\sqrt{T}$ to $1/T$ power-law crossover in thermalized systems, a behavior also reported for a finite-size nonlinear unitary-circuit lattice
model~\cite{Merab22}, suggesting it may be a generic feature of thermalization in nonlinear lattices. The probability distribution of local amplitudes $p(A)=y_0^2(A)$ provides an additional confirmation: in the Gibbs regime for $D=0.25$, the numerically obtained $p(A)$ agrees closely with the TIO prediction, confirming thermal equilibrium. As the energy density increases, the tail of $p(A)$ progressively deviates toward heavier forms, directly reflecting the growing role of large-amplitude localized excitations in the non-Gibbs and weakly nonergodic regimes. 

{We defined the crossover to the weakly nonergodic regime by the onset of a divergent mean excursion time, as indicated by the power-law tail of the excursion-time PDF.} {In the weakly nonergodic regime, we observed long-lived two-site localization for $D>1$, whereas $D\leq1$ favors one-site localization.}
{More generally, we characterized this localization through the high-energy compacton branches, associating the dynamically relevant structures with energy maximizers that concentrate excess energy into one or a few effective lattice nodes, in analogy with the energy minimizers of~\cite{MarzuolaMattingly2025}.} {Interestingly, three-site compactons initialized at $t=0$ are found to robustly survive for very long times, e.g.\ $\sim10^7$, only in the weakly nonergodic regime, providing a dynamical counterpart to the fixed-mass compact state identified therein.}

The above analysis {raises} a variety of questions regarding the thermodynamic analysis of such nonlinear dynamical lattices. Perhaps foremost among them is
whether a modified statistical description can be established for the apparently ergodic yet non-Gibbsian regime, both here and in similar models~\cite{MithunDNLS18}. On a related note, the present work characterizes the weakly nonergodic regime through energy maximizers; this is a regime that certainly merits further
study. More generally, the statistical mechanics of {nonlinear} lattices remains an intriguing nexus of different fields, where the potential inclusion of more conservation laws, {when present,} or the analysis of higher-dimensional systems, {when physically relevant,} as, e.g., for the
DNLS~\cite{kev09}, are bound to provide numerous future insights.

\section*{Acknowledgments}
This work was funded by the Deutsche Forschungsgemeinschaft (DFG) – Project-ID 414984028 – SFB 1404 FONDA and supported through the DFG Walter Benjamin Programme (Grant No.~506198590) (Y.K.). 
We acknowledge support from the Ministry of Science, Technological Development and Innovation of the Republic of Serbia, Grant No.~451-03-33/2026-03/200017,
~451-03-34/2026-03/200124. 
This material is based upon work supported by the US National Science Foundation under Grant No. PHY-2110030, PHY-2408988 and DMS-2204702 (P.G.K.). 
This research was partly conducted while P.G.K. was 
visiting the Okinawa Institute of Science and Technology (OIST) through the Theoretical Sciences Visiting Program (TSVP) and the Sydney Mathematical Research
Institute (SMRI) of the University of Sydney. This work was also supported by the Simons Foundation [SFI-MPS-SFM-00011048, P.G.K.].
The authors also acknowledge constructive discussions with Professor 
Jeremy Marzuola in connection to the work of~\cite{MarzuolaMattingly2025}
and its connections to the present manuscript.

\section*{Data Availability}
The numerical datasets used to generate the figures in this work are deposited 
on Zenodo~\cite{kati_data_allv} and the source codes are publicly available on 
the edoc-Server of Humboldt-Universität zu Berlin~\cite{Kati2026code}; an extended 
version with additional integration methods is available on 
GitHub~\cite{Kati2026github}. 
All are currently under restricted access. Access prior to public release is 
available from the corresponding author upon reasonable request.
\appendix

\section{Phase Diagram for Toy Model}\label{sec:temperaturecurves}

\subsection*{Infinite temperature, \texorpdfstring{$\beta\to 0$}{beta -> 0}}
\label{sec:toy_beta0}

We consider the infinite-temperature limit $\beta\to 0$ with
\begin{equation}
\mu:=\beta\alpha \quad \text{fixed}\ {(\mu>0)}, \qquad (\alpha\to\infty).
\end{equation}
{In this limit, the Bessel kernel in the partition function of
Eq.~(\ref{partitionfunction}) reduces to $I_0(0)=1$, while the Gaussian
factor $e^{-(\beta/4)A_\ell^2}$ tends to unity. 
Hence the partition function factorizes over sites,
\begin{equation}
Z \simeq (2\pi)^N \prod_{\ell=1}^N
\int_0^\infty dA_\ell\, e^{-\mu A_\ell}.
\end{equation}
Consequently, the normalized single-site distribution of the amplitude
$A=|b_\ell|^2$ is
\begin{equation}
\rho(A)=
\frac{e^{-\mu A}}{\int_0^\infty e^{-\mu A}\,dA}
=\mu e^{-\mu A},\qquad A\in[0,\infty).
\end{equation}
The required moments are 
\begin{equation}
\langle A^n\rangle
=\mu\int_0^\infty A^n e^{-\mu A}\,dA
=\frac{n!}{\mu^n},
\qquad n=0,1,2 .
\end{equation}
Therefore, the norm density is
\begin{equation}
a=\langle A\rangle=\frac{1}{\mu}.
\end{equation}
In the limit $\beta \rightarrow 0$ with $\mu=\beta\alpha$ fixed, the sites are decoupled and the phases are independent and
uniformly distributed. The phase-dependent interaction term in
Eq.~(\ref{hamiltoniantoy}) therefore averages to zero, so the energy
density is determined only by the onsite quartic term:
\begin{equation}
h=\frac{1}{4}\langle A^2\rangle
  =\frac{1}{4}\frac{2}{\mu^2}
  =\frac{1}{2\mu^2}.
\end{equation}
Eliminating $\mu$ using $a=1/\mu$ gives the infinite-temperature curve
\begin{equation}
h(\beta=0)=\frac{a^2}{2}.
\end{equation}
}

\subsection*{Zero temperature (\texorpdfstring{$\beta\to\infty$}{beta -> infinity})}

In the zero-temperature limit ($\beta\to\infty$), the lowest-energy homogeneous-norm configuration ($A_\ell=a$) is obtained for a staggered phase pattern with $\Delta\theta=\pi$, i.e., $\theta_\ell=\ell\pi$. Then
\begin{equation}\label{eq:bell_state}
b_\ell=\sqrt{a}\,e^{i\ell\pi}, \qquad |b_\ell|^4=a^2,
\end{equation}
and
\begin{equation}\label{eq:bell_coupling}
\begin{aligned}
{b_\ell^*}^2 b_{\ell-1}^2 + b_\ell^2 {b_{\ell-1}^*}^2
&=2a^2\cos\!\big(2(\theta_\ell-\theta_{\ell-1})\big)\\
&=2a^2\cos(2\pi)=2a^2 .
\end{aligned}
\end{equation}
Substituting Eqs.~(\ref{eq:bell_state})--(\ref{eq:bell_coupling}) into Eq.~(\ref{hamiltoniantoy}) yields the corresponding energy density
\begin{equation}\label{eq:minenergy}
h=\frac{H}{N}
=\frac{a^2}{4}-\frac{D}{4}\,2a^2
=\frac{a^2}{4}(1-2D).
\end{equation}

\bibliographystyle{apsrev4-2}
\bibliography{bibl}

\end{document}